\documentclass[12pt]{article}
\usepackage{latexsym}
\usepackage{amsmath,amsbsy,amssymb}
\usepackage{bm}
\usepackage{verbatim}
\usepackage{graphicx}

\usepackage{cite}

\let\OLDthebibliography\thebibliography
\renewcommand\thebibliography[1]{
  \OLDthebibliography{#1}
  \setlength{\parskip}{0pt}
  \setlength{\itemsep}{0pt plus 0.3ex}
}

\newcommand{\dd}{\mbox{\rm d}}
\newcommand{\wg}{\wedge}
\newcommand{\gam}{\gamma}
\newcommand{\Gam}{\Gamma}
\newcommand{\dg}{\dagger}

\newcommand{\tl}{\tilde}
\newcommand{\ul}{\underline}
\newcommand{\om}{\omega}
\newcommand{\lam}{\lambda}

\newcommand{\dotx}{{\dot x}}

\newcommand{\DD}{\mbox{\rm D}}

\newcommand{\p}{\partial}
\newcommand{\be}{\begin{equation}}
\newcommand{\bear}{\begin{eqnarray}}
\newcommand{\ear}{\end{eqnarray}}
\newcommand{\ee}{\end{equation}}

\newcommand{\lbl}{\label}
\newcommand{\bi}{\bibitem}
\newcommand{\ci}{\cite}

\newcommand{\vs}{\vspace}
\newcommand{\hs}{\hspace}

\newcommand{\vphi}{\varphi}
\newcommand{\sg}{\sigma}
\newcommand{\vac}{|0 \rangle}
\newcommand{\ba}{{\bar a}}
\newcommand{\bb}{{\bar b}}

\newcommand{\bp}{{\bm p}}

\newcommand{\bx}{{\bm x}}

\newcommand{\hp}{{\hat p}}
\newcommand{\al}{{\alpha}}
\newcommand{\bet}{{\beta}}

\begin{document}

\

\baselineskip .7cm 

\vs{8mm}

\begin{center}
	
	{\LARGE \bf On Negative Energies, Strings, Branes, and Braneworlds:
		A Review of Novel Approaches}

\vs{3mm}

Matej Pav\v si\v c

Jo\v zef Stefan Institute, Jamova 39,
1000 Ljubljana, Slovenia

e-mail: matej.pavsic@ijs.si

\vs{6mm}

{\bf Abstract}
\end{center}

\baselineskip .4cm 

{\footnotesize
On the way towards quantum gravity and the unification of interaction, several ideas have
been rejected and avenues avoided because they were perceived as physically unviable. But
in the literature there are works in which it was found the contrary, namely
that those rejected topics make sense after all. Such topics, reviewed in this article, are negative energies occurring
in higher derivative theories and ultrahyperbolic spaces, ordering ambiguity of operators
in curved spaces, the vast landscape of possible compactifications of extra dimensions in
string theory, and quantization of a 3-brane in braneworld
scenarios.

\vs{2mm}


\vs{1mm}

}

\baselineskip .6cm

\section{Introduction}

After eighty years quantum gravity is still unfinished project. If starting with
the Einstein-Hilbert action, it turns out that the theory is non renormalizable,
unless one introduces higher derivative terms of the form $R^2$. But then
negative energies enter the description. It has been taken for granted that a
system in which negative energies are admissible cannot be stable.
But in Refs.\,\ci{Smilga:2004cy,Smilga:2005gb,Smilga:2008pr,Robert:2006nj,
Pavsic:2013noa,Pavsic:2013mja,Pavsic:2016ykq} it has been shown that this is not
necessarily the case. Smilga found that there are islands of stability where
even the interacting Pais-Uhlenbeck oscillator, a toy model for higher derivative
theories, behave stably. In Ref.\,\ci{Pavsic:2013noa,Pavsic:2013mja,Pavsic:2016ykq} it was pointed out that
instability occurs if the interaction potential is unbounded, whilst in the
presence of a physically realistic potential which is bounded from below and
from above, no instabilities occur, even if the system possesses negative energy
states. This was further elaborated in the book\,\ci{PavsicStumbling}, where
many scenarios have been investigated from the scattering of a particle on
a fixed potential to the collisions of two particles. Numerical calculations
reveal fascinating behavior, a notable feature being that after
scattering or collision particles can have significantly higher absolute value
of kinetic energy than initially.

In the same book\,\ci{PavsicStumbling} also many other problematic topics
related to quatnum gravity and unification of forces and particles have
been considered and clarified. One subtle problem is ordering ambiguity of
quantum operators in curved spaces. In Ref.\,\ci{PavsicOrder} it was shown
how the usage of the geometric momentum operator ${\hat p} = - i \gam^a \p_a$,
where $\gam^a$ are position dependent basis vectors, gives an unambigous
expression for the squared operator ${\hat p}^2$ acting on a scalar field.
In Ref.\,\ci{PavsicStumbling} it was shown that if ${\hat p}^2$ acts on a spinor field, $\psi$,
then the expression ${\hat p}^2 \psi$ so obtained is also unambigous and, besides
$\DD_\mu \DD^\mu \psi$, it contains a term with the Riemann tensor coupled 
to the spin tensor.

Another way to quantum gravity is via strings. In Ref.\ci{PavsicSaasFee} it
was shown that a string 
can be consistently quantized in a space with equal number of space-like and
time-like dimensions, e.g., in a space with neutral signature. A
string extended along one intrinsic dimension is a particular case of a brane, i.e., an
object extended along arbitrary number of intrinsic dimensions. 
An attractive possibility is to consider our universe as a brane living in a higher
dimensional space. Such braneworld scenarious are very popular subjcet of
investigation (see. e.g., \ci{Rubakov,Akama,GibbonsBrane,PavsicEmbed1,PavsicEmbed2,PavsicEmbed3,PavsicEmbed4,Gogberashvili1998,PavsicEmbed5,Gogberashvili2000,PavsicTapia,Brax,PavsicBook}). Quantum gravity could be achieved
if we were able to quantize the brane, which is very difficult, unless one takes a larger
view and introduces the concept of flat brane, which is just a bunch of free particles that upon quantization are described by a continuous set of non interacting quantum
fields\ci{PavsicBrane,PavsicStumbling}. Introducing interactions among those fields, one
obtains classical branes as expectation values.

In the following sections I will review the topics mentioned above. In Sec.\,2 I will describe behavior of classical and then quantum systems with both positive and negative energy degrees of freedom. Then I will present a resolution of ordering ambiguities (Sec.\,3).
Strings and branes from the novel point of view will be considered in Sec.\,4.

\section{Behavior of systems with negative energies}

\subsection{Classical theory}

When I was ten year schoolboy, our geography teacher brought into the class
a globe and told us that it represents the Earth. Some pupils wondered how
is it possible that the people in Australia do not fall down. They took for
granted that the ``falling down'' direction is from up to down, and hence the
Australian should fall down and not stay on the Earth. Such a naive view has
resemblance in the wide spread believe that the presence of negative energy
states in a system implies its never ending ``falling down'' behavior (e.g., rolling down a potential unbounded from below). It is true that under the
influence of a potential unbounded from below, a particle with a {\it positive}
kinetic energy rols down the potential. However, a particle with a {\it negative}
kinetic energy does not role down such potential but climbs up, and is stable
on the top of the potential. The opposite holds for a potential that
is unbounded from above.

As a model let us consider a system with two degrees of freedom, described
by the Lagrangian
 \be
L = \frac{m}{2}({\dot x}^2 - {\dot y}^2) - V(x,y),
\lbl{2.1}
\ee
The equation of motion,
 \be
m{\ddot x} = - \frac{\p V}{\p x} ,
\lbl{2.2}
\ee
\be
m{\ddot y} = \frac{\p V}{\p y},
\lbl{2.3}
\ee
show that the degree of freedom $x$ with positive kinetic energy experiend the
force $F_x = - \frac{\p V}{\p x}$, whilst the degree of freedom $y$ with
negative kinetic energy experiences the force $F_y = \frac{\p V}{\p y}$.

Hence, if $V(x,y)= \frac{k}{2}(x^2 + y^2)$, $k>0$ the system falls down the potential
in the $x$-direction, and climbs up the potential in the $y$-direction.
Such system is thus not stable, though the potential is bounded from below, it manifests a runaway behavior of the $y$ degree of freedom. Analogous holds
for the potential $V(x,y)= -\frac{k}{2}(x^2 + y^2)$, $k>0$, only the roles
of $x$ and $y$ are interchanged, so that the positive energy degree of freedom
$x$ exhibits runaway behavior, whilst the negative energy degree of freedom
$y$ behaves stably.

However, if $V(x,y)= \frac{k}{2} (x^2-y^2)$, then both degrees of freedom,
$x$ and $y$, behave stably. Their equations of motion are
\bear
  &&m \ddot x + k x =0,  \lbl{2.4}\\
  &&m \ddot y + k y =0.  \lbl{2.5}
\ear

According to my experience, such an explicit, lengthy, explanation is necessary,
because a brief explanation was often not understood correctly. A typical
objection was that a system with a potential unbounded from below cannot
be stable. People somehow overlooked the points discussed above, because
they were so much accustomed to thinking that a potential must have a minimum
(at least a local one), otherwise the system is unstable.

However, also among the experts who are aware of the points discussed
through Eqs.\,\ref{2.1}--\ref{2.5}, there is consensus that the presence of
the degrees of freedom with negative energies implies instability anyway,
even if the potential is like  $V(x,y)= \frac{k}{2}(x^2 - y^2)$. Namely,
if there is a coupling between positive and negative energy degrees of
freedom, then energy flows between them and causes a runaway behavior of the
system. In Refs.\,\ci{Pavsic:2013noa,Pavsic:2016ykq,PavsicStumbling} it was shown that such runaway behavior indeed happens ---
in the presence of physically unrealistic potentials.

An example of physically unrealistic potential is
\be
V= \frac{k}{2} (x^2 -y^2) + \frac{\lambda}{4}(x^2-y^2)^2 .
\lbl{2.6}
\ee
It is unrealistic, because it is unbounded from above due to the presence of
the quartic term which prevail at larger values of $x$ and $y$. The numerical
calculation, performed in Refs.\,\ci{Pavsic:2013noa,Pavsic:2016ykq,PavsicStumbling}
show that such a system is indeed unstable.
It exhibits ever increasing oscillations, both in the $x$ and $y$ direction.
The kinetic energy oscillates with for ever increasing amplitudes (Fig.\,1). The total
energy $E_{\rm tot} = \frac{m}{2} ({\dot x}^2-{\dot y}^3) + V(x,y)$ remains
constant (within the numerical error), as it should, because the Lagrangian is
invariant with respect to translations in time.

\setlength{\unitlength}{.68mm}
\begin{figure}[h]

	\centerline{\includegraphics[scale=0.45]{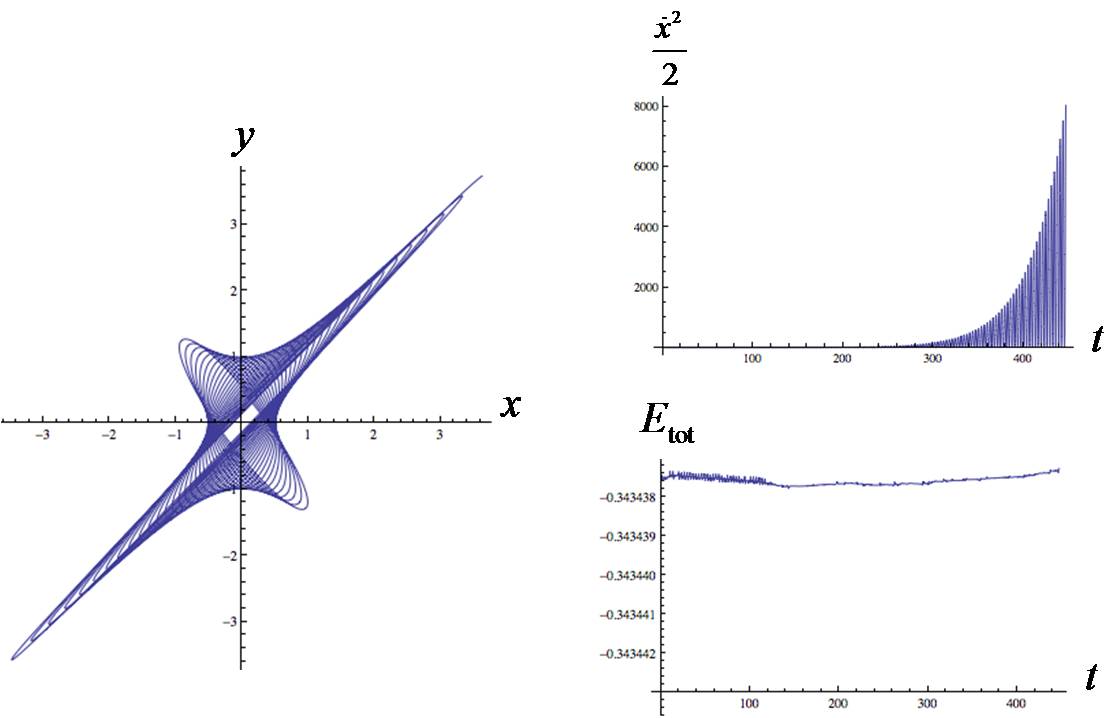}}

	\caption{\footnotesize The solution of the system
		described by Eqs.\, (\ref{2.2}),(\ref{2.3}), with $m=1$, in the
		presence of the unbounded potential (\ref{2.6}) with $k=1$, $\lambda=0.1$, for the initial conditions
		${\dot x} (0)=1$, ${\dot y} (0)=-1.2$, $x(0)=0$, $y(0)=0.5$.}
	\lbl{fig2.2}
\end{figure}

If instead of the unbounded potential (\ref{2.6}) we take a bounded potential,
for instance
	\be
V = {\rm e}^{-a(x^2+y^2)} \left [ \frac{1}{2} (x^2-y^2)+
\frac{\lambda}{4} (x^2-y^2)^2\right ].
\lbl{2.7}
\ee
\setlength{\unitlength}{.72mm}
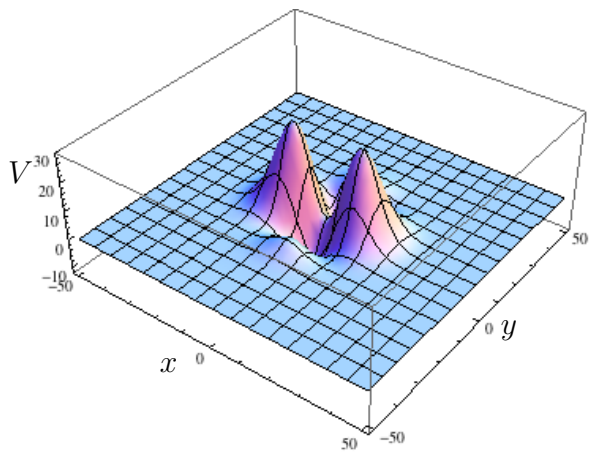
\begin{figure}[h]

	\hs{11mm}\begin{picture}(120,76)(0,0)
	
	\put(25,-5){\includegraphics[scale=0.594]{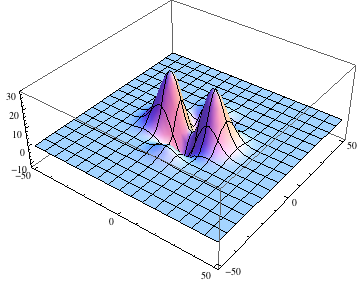}}
	
	\put(22,45){$V$}
	
	\put(50,10){$x$}
	\put(113,17){$y$}
	
	
	\end{picture}
	
	\caption{\footnotesize An example of the potential that is bounded from below and from above (see Eq.\,(\ref{2.7})).}
	\lbl{fig2.2}
\end{figure} 	
as shown in Fig.\,2, then we obtain stable behavior\,\ci{Pavsic:2013noa,Pavsic:2016ykq,PavsicStumbling}. In Fig.\,3\,(up) are shown the results
of numerical calculation of the equation of motion (\ref{2.2}),(\ref{2.3}),(\ref{2.7}) for the same parameters and the initial
data as in Fig.\,1, and for the damping constant $a=0.001$ in the potential
(\ref{2.7}). Now  the system, instead of escaping into infinity, oscillates
within a finite region of the $(x,y)$-space. The amplitude of energy oscillations
does not increas to infinity, but stays within an oscillating envelop. 
\setlength{\unitlength}{.68mm}

\begin{figure}[h!]

	\centerline{\includegraphics[scale=0.45]{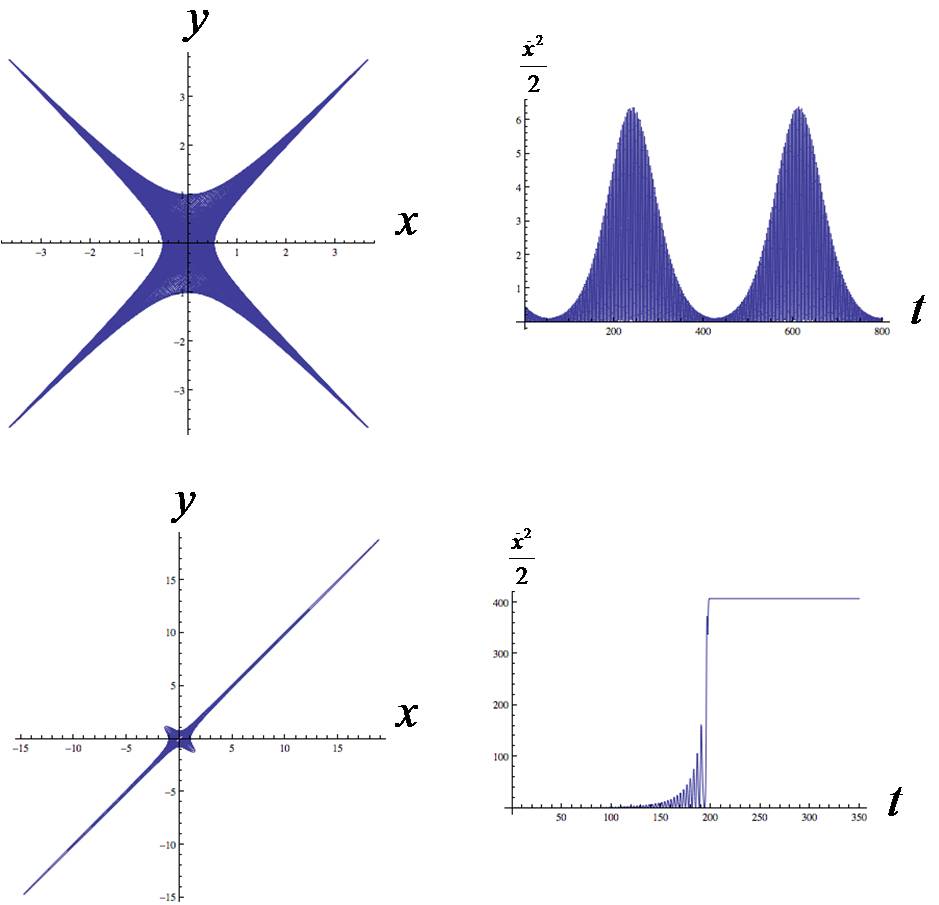}}
	
	\caption{\footnotesize The solution of the system
		described by Eqs.\,(\ref{2.2}),(\ref{2.3}), with $m=1$, and Eq.\,(\ref{2.7})
		with $k=1$, $\lambda=0.1$, $a=0.001$, for the initial conditions
		${\dot x} (0)=1$, ${\dot y} (0)=-1.2$, $x(0)=0$, $y(0)=0.5$ (up). The same system, except that the initial velocity is increased from ${\dot x}(0)=1$ to ${\dot x}(0)=1.5$ (down).}
	\lbl{fig2.4}
\end{figure}

If we increase the initial velocity, then the system, after oscillating
for some time within a finite region of the $(x,y)$-plane, escapes towards
infinity with a finite constant velocity, as shown in Fig.\,3\,(down). With time, the amplitude of the velocity oscillations increases until the velocity reaches a finite constant value. Such system is thus stable in the sense that its
velocity remains finite, and does not approach to infinity as in the case of
the unbounded potential (\ref{2.6}).

In Ref.\,\ci{PavsicStumbling}, numerical calculation were performed also
for the potential 
\be
V = \frac{k}{2} ({\rm sin^2\, x -{\rm sin}^2 \,y )
	+\lambda {\rm sin}\,x \,{\rm sin} y}.
\lbl{2.8}
\ee
which extends into infinity in the $(x,y)$-plane (Fig.\,4),
and for the potential
\be
V = \left ( \frac{k}{2} ({\rm sin^2\, x -{\rm sin}^2 \,y )
	+\lambda {\rm sin}\,x \,{\rm sin} y} \right )  \left (1 + {\rm tanh} (r_0 - \sqrt{x^2 + y^2}) \right  )
\lbl{2.9}
\ee
which vanishes outside the domain determined by $r_0$ (Fig.\,5).
 \setlength{\unitlength}{.5832mm}
\begin{figure}[h!]
	
	\centerline{\includegraphics[scale=0.45]{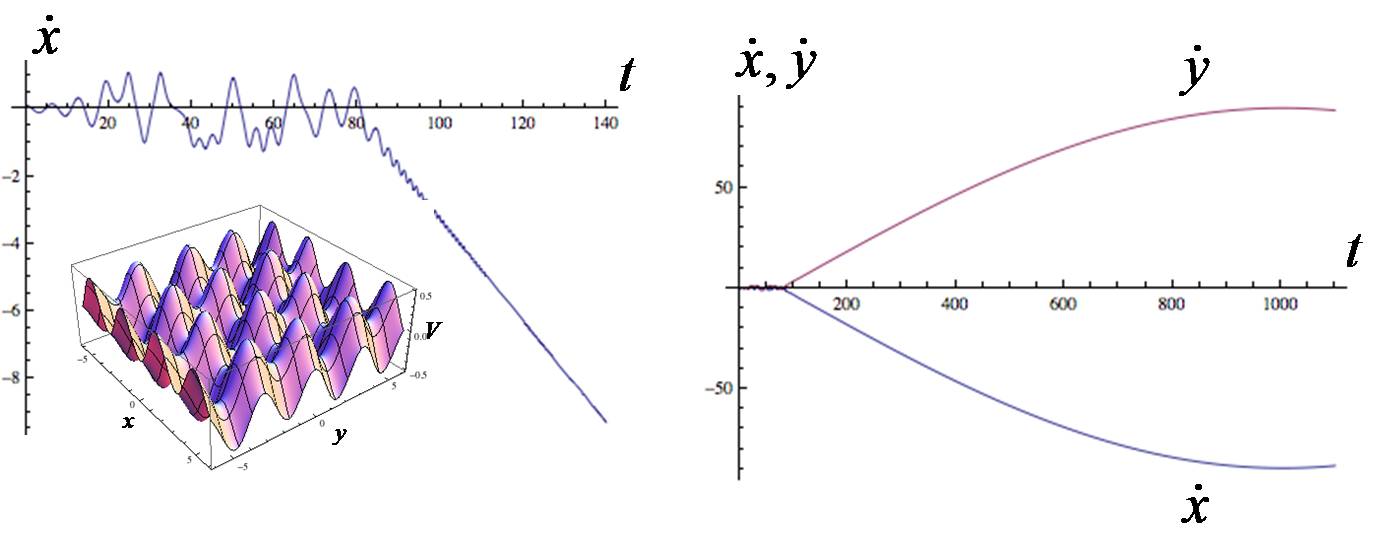}}
	
	\caption{\footnotesize Velocity ${\dot x}$ as function of time for the same initial data as in Fig.3.
		In the left plot the time interval is from
		$t=0$ to $t=140$. In the right plot the time interval is extended to $t=1200$, and the upper curve
		belongs to ${\dot y}$. Potential is that of Eq.\,(\ref{2.8}).}
\end{figure}
\begin{figure}[h!]
	
	\centerline{\includegraphics[scale=0.45]{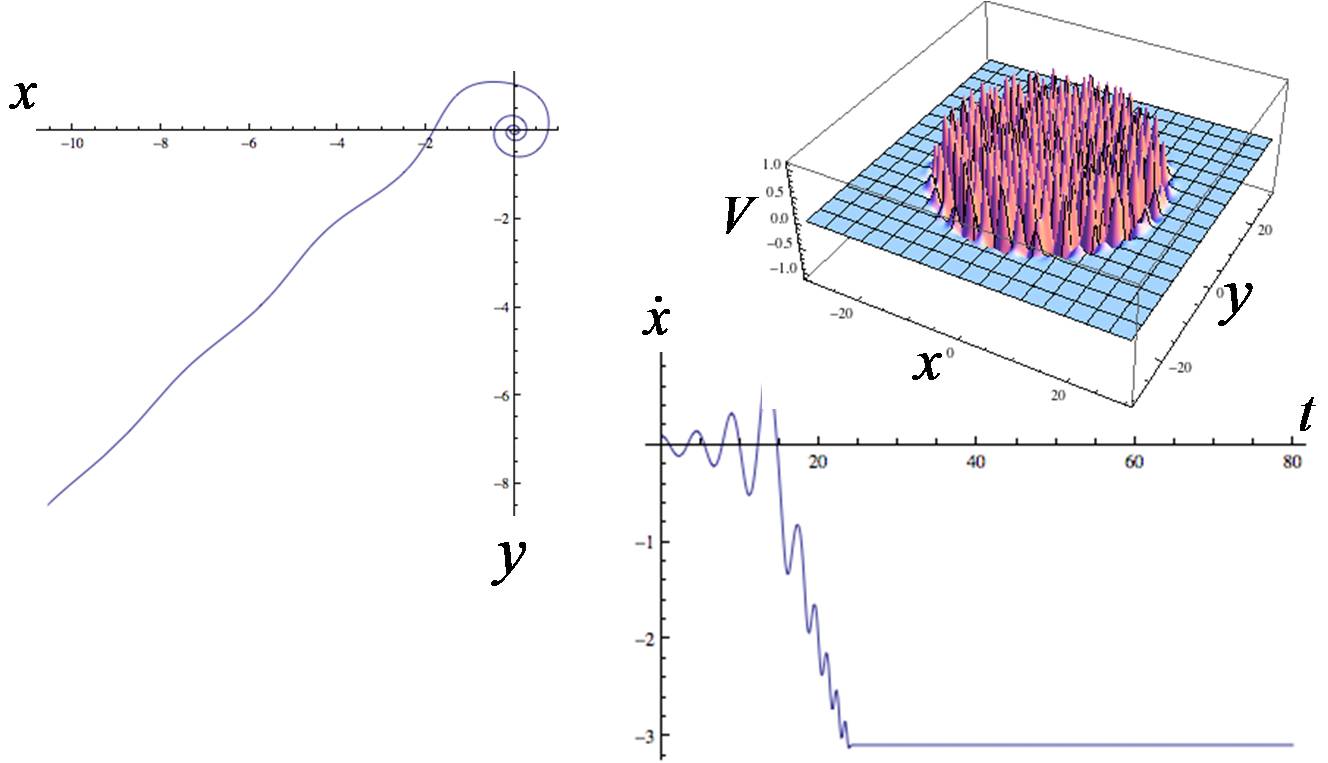}}
	
	\caption{\footnotesize  Potential cut at $r_0=20$ ($\lambda=0.3$, ${\dot x}=0.1$,
		${\dot y}=0$, $x(0)=0.0000001$, $y(0)=0$). The system escapes with a constant velocity. }
\end{figure} 

Motion of a particle described by the Lagrangian (\ref{2.1}) with the potentials
(\ref{2.8}) and (\ref{2.9}) exhibits peculiar behavior as described in
Ref.\ci{PavsicStumbling}. Two examples are shown in Fig.\,4 and 5.
Under certain initial conditions the particle can perform huge jumps in
the $({\dot x},{\dot y})$-space. The particle may start with a low initial
velocity and acquire a very high velocity.

In all those cases with bounded potentials a runaway behavior of the system's velocity does not occur. The velocity and hence the corresponding component
of kinetic energy remains finite. If during the evolution of the system
there is an increase of the absolute value of the kinetic energy, it cannot
exceed the overall potential difference---if the potential is assymptocially
constant. This is so because the total energy is constant:
\be
   E_x + E_y + V = C,
\lbl{2.9a}
\ee
where  $E_x = {\dot x}^2/2$, $E_y=-{\dot y}^2/2$ and $C$ a constant. Therefore,
\be
    E_x(t)+E_y(t)= E_x(0)+E_y(0)+ V(t)-V(0),
\lbl{2.10}
\ee
where $V(t)$ is the value of the potential experienced by the particle at time
$t$. In the case of a bounded potential, Eq.\,(\ref{2.10}) implies that
$ E_x(t)+E_y(t)$ is finite if $E_x(0)+E_y(0)$ is finite.

Having clarified the behavior of the system based on the Lagrangian (\ref{2.1})
for various potentials, let us now consider a more general Lagrangian,
  \be
  L = \frac{1}{2} g_{\alpha \beta} {\dot x}^\alpha {\dot x}^\beta -
  \frac{1}{2} k_{\alpha \beta}x^\alpha x^\beta
  - \frac{\lambda}{4} \lambda_{\alpha \beta \gam \delta} x^\alpha x^\beta x^\gamma x^\delta ,
  \lbl{II2.39}
  \ee
 a special
 case of which is the Lagrangian (\ref{2.1}).
  For a particular choice of $g_{\alpha \beta}$, $k_{\alpha \beta}$ and
  $\lambda_{\alpha \beta \gamma \delta}$, we obtain
  \be
  L = \frac{1}{2} \left (m_x {\dot x}^2 - m_y {\dot y}^2 \right )
  -\frac{k}{2}(x^2-y^2)- \frac{\lambda}{4}(x^2-y^2)^2,
  \lbl{II2.40}
  \ee
  where in general $m_x \neq m_y$ $k_x =k_y = k$.
  
  Concerning other possibilities, in Ref.\,\ci{PavsicStumbling} we find:
  
\begin{quote}  
  Another possibility is such a choice of $g_{\alpha \beta}$, $k_{\alpha \beta}$ and
  $\lambda_{\alpha \beta \gamma \delta}$ which gives
  \be
  L = \frac{1}{2} \left ({\dot x}^2 - {\dot y}^2 \right )
  -\frac{1}{2}(\om_x x^2- \om_y y^2)- \frac{\lambda}{4}(x+y)^4,
  \lbl{II2.41}
  \ee 
  with $m_x = m_y=1$, $k_x \neq k_y$, and an appropriate choice of the coefficients
  $\lambda_{\alpha \beta \gam \delta}$, i.e., $\lam_{x x x x}= \lam$,
  $\lam_{xxxy} = \lam_{xxyx} = \lam_{x y x x} = \lam_{y x x x} = \lam/4$, etc.\..
  We now write $k_x = \om_1^2$, $k_y = \om_2^2$.
  
  Introducing the new variables
  \be
  u= \frac{x+y}{\sqrt{2}}~,~~~~v= \frac{x-y}{\sqrt{2}},
  \lbl{II2.42}
  \ee
  the Lagrangian (\ref{II2.41}) becomes
  \be
  L= {\dot u}{\dot v}- \frac{1}{4} \left [ (\om_1^2 - \om_2^2) (u^2 + v^2)+
  2 (\om_1^2 + \om_2^2) u v  \right ] - \lam u^4 .
  \lbl{II42}
  \ee
  The equations of motion are
  \be
  {\ddot u} +  \frac{1}{2} (\om_1^2 + \om_2^2) u + \frac{1}{2} (\om_1^2 - \om_2^2) v
  =0,
  \lbl{II2.44}
  \ee
  \be
  {\ddot v} +  \frac{1}{2} (\om_1^2 + \om_2^2) v + \frac{1}{2} (\om_1^2 - \om_2^2) u
  + 4 \lam u^3 =0,
  \lbl{II2.45}
  \ee
  Eliminating $v$ from the first equation and inserting it into the second equation,
  we obtain
  \be
  u^{(4)}+(\om_1^2 + \om_2^2){\ddot u} + \om_1^2 \om_2^2 u - \Lambda u^3 = 0,
  \lbl{II2.46}
  \ee
  where $\Lambda = 2 (\om_1^2 - \om_2^2)\lam$. 
  We have thus obtained the forth order equation ($u^{(4)}\equiv \stackrel{....}{ u}$),
  known as the self-interacting Pais-Uhlenbeck oscillator, which can be written as
  \be
  \left ( \dfrac{\dd^2}{\dd t^2}+ \om_1^2 \right )
  \left ( \dfrac{\dd^2}{\dd t^2}+ \om_2^2 \right )u - \Lambda u^3 = 0.
  \lbl{II2.47}
  \ee
  The corresponding Lagrangian is
  \be
  L= \frac{1}{2} \left [ {\ddot u}^2 - (\om_1^2 + \om_2^2){\dot u}^2 +
  \om_1^2 \om_2^2 u^2 	 \right ] + \frac{\Lambda}{4}u^4 .
  \lbl{II2.48}
  \ee 
  The action is given by the integral of the above Lagrangian over time, and after omitting the surface term, we have
  \be
  I = \int \dd t \left [ \frac{1}{2}u \left ( \dfrac{\dd^2}{\dd t^2}+ \om_1^2 \right )
  \left ( \dfrac{\dd^2}{\dd t^2}+ \om_2^2 \right )u 
  + \frac{\Lambda}{4}u^4 \right ] .
  \lbl{II2.49}
  \ee
  Pais-Uhlenbeck oscillator is a toy model for higher derivative field theories of
  the form
  \be
  I = \int \dd^4 x\left [ \frac{1}{2} \phi \left (\Box + m_1^2 \right )
  \left ( \Box + m_2^2 \right )\phi 
  + \frac{\Lambda}{4}\phi^4 \right ]
  \lbl{II2.50}
  \ee
  in which the spatial dependence of the field $\phi (x)$, $x \equiv x^\mu =
  (t, \bx)$, is supressed. Higher derivative theories, including $R+R^2$
  gravity, are usually considered as problematic, because they include negative
  energies.
  We have seen that they are not problematic if there is no interaction
  term, such as $\frac{1}{4} \Lambda u^4$ in eq.\,(\ref{II2.48}), or $\frac{\lambda}{4}(x^2-y^2)^2$ in Eqs.\,(\ref{II2.49}),(\ref{II2.50}). Then the
  system behaves like a system of free oscillators, and it does not matter
  whether some of them have positive and some negative energies. Such a
  free system can be described by a Lagrangian in which all degrees of freedom
  have positive energies.
  Several authors\,\ci{Mostafazadeh:2010yw,Banerjee:2013upa,Bolonek,Sokolov,
  	Bolonek:2006ir,Bagarello,Nucci,Masterov:2015ija,Masterov:2016jft,Dector,Ghosh}
  have found that the (free) Pais-Uhlenbeck oscillator
   can be described in terms of positive energies only.
  This is no longer so in the presence of an interaction. The Lagrangian (\ref{II2.48})
  cannot be transformed into a Lagrangian for two mutually interacting positive energy
  oscillators\,\ci{Pavsic:2013noa,Pavsic:2016ykq}. One of them necessarily has negative energy.
  It has been generally believed that
  such a system is unstable. But numerical calculations show that it is not
  necessarily so\,\ci{Smilga:2004cy,Smilga:2005gb,Smilga:2008pr,Robert:2006nj,Pavsic:2013noa}. The system is stable for certain range
  of the initial velocity and the coupling constant $\Lambda$. But in Ref.\,\ci{Robert:2006nj}
  an unconditionally stable interacting system was found.
  
  If instead of the Lagrangian (\ref{II2.48}) with the unbounded interaction potential
  $\frac{1}{4} \Lambda u^4$ we take a bounded potential, such as $\frac{1}{4} \Lambda {\rm sin}^4 u$, or $\frac{1}{4} \Lambda {\rm e}^{-(x+y)^2}$, then numerical calculation show\,\ci{Pavsic:2013noa}
  that such system is stable in the variable $u$, whilst in the
  variables $x$, $y$ is stable in the sense that their time derivatives (velocities)
  cannot escape into infinity, but approach finite values, whereas the $x$ and $y$ asymptotically behave as coordinates of a free particle and increase linearly with time.
\end{quote}

Exhaustive treatment of higher derivative theories and their stability can be found
in Refs.\cite{Kaparulin:2020rqz,Abakumova:2019dov,Abakumova:2019wpn,Kaparulin:2019njc,
Abakumova:2019ifi,Kaparulin:2018npv,Abakumova:2017uto}.

Besides scattering of a particle on a fixed potential one can also consider
collisions of particles. A toy model, considered in Ref.\ci{PavsicStumbling},
was determined by the Lagrangian
 \be
L = \frac{1}{2} \left ( {\dot x}_1^2 - {\dot y}_1^2 + {\dot x}_2^2 - {\dot y}_2^2 \right )
- V(x_1-x_2, y_1-y_2),
\lbl{2.11}
\ee
for the interaction potential
   $$  V = {\rm e}^{-a [(x_1-x_2)^2+(y_1-y_2)^2]}
\left  [ \frac {1}{2} [(x_1-x_2)^2-(y_1-y_2)^2] \right . \hs{2cm}$$
\be  
\hs{3cm}+ \left . \frac{k}{4}[(x_1-x_2)^2-(y_1-y_2)^2] \right ] .
\lbl{2.12} 
\ee
Numerical solution of the corresponding equations of motion
exhibit very interesting behavior\ci{PavsicStumbling}. Namely, the velocities
of the incoming particles can be small, whilst the velocities of the
outgoing particles can increase by a factor which is about ten. Again,
the total energy in the process was conserved, and the increase of absolute
values of the final kinetic energies was due to the potential difference.
Because the interaction potential was bounded, the particles escaped with
a constant finite velocity.

Another case, considered in Ref.\ci{PavsicStumbling}, was given by
the Lagrangian
\be
\frac{m_1}{2} \left ( {\dot x}_1^2 + {\dot y}_1^2  \right ) + 	\frac{m_2}{2} \left ( {\dot x}_2^2 + {\dot y}_2^2 \right )
- V(x_1-x_2, y_1-y_2).
\lbl{2.13}
\ee
with
 \be
V(x_1-x_2, y_1-y_2) = - \frac{m_1 m_2}{\sqrt{(x_1-x_2)^2+(y_1-y_2)^2}}
\lbl{2.14}
\ee
This model describes two gravitationally interacting particles, with the
third dimension suppressed.
\begin{figure}[h!]

	\centerline{\includegraphics[scale=0.55]{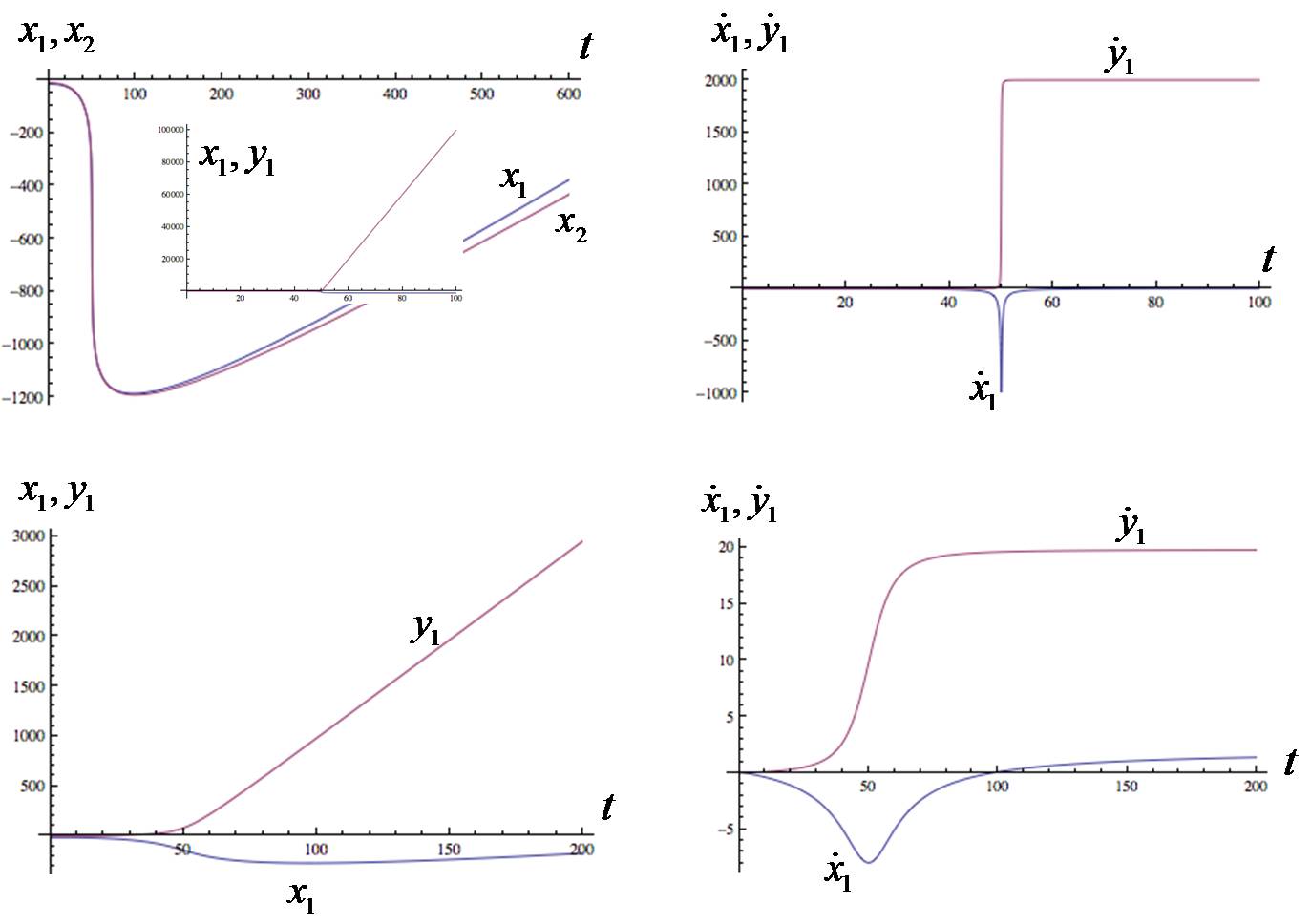}}
	
	\caption{\footnotesize  A slight deviation from a head on collision brings no runaway behavior of velocity.
		Up: $x_1(0)=-15$, $y_1(0) =0.01$, $x_2 (0) = -10$, $y_2(0)=0$. Down: $x_1(0)=-15$, $y_1(0) =1$, $x_2 (0) = -10$, $y_2(0)=0$
		In both cases the initial velocities are the same: $x'_1(0)=0.1, y'_1(0)=x'_2(0)=y'_2(0)=0$, and
		$m_1=1$ and $m_2=-1$}
	\lbl{fig2.15}
\end{figure}

It is well known\,\ci{Bondi,Shanshan,Sheng,Mei1,Mei2,Choi1,Choi2,Choi3} that if $m_2=-m_1 <0$, then the so called
diametric drive takes place: both particles move into the same direction so that
their velocities increase with constant acceleration. In Ref.\,\ci{PavsicStumbling} its was shown by numerical calculations that such
behavior occur only if we take exactly equal absolute values of masses and exactly
zero relative velocity, which is an unrealistic situation. In reality the
masses cannot be exactly the same (apart from the sign) and the relative velocity cannot be exactly zero. Then, however small is the difference $|m_1|-|m_2|\neq0$, or
however small, but finite, is the relative vellocity, the runaway behavior does not go for ever. After a certain
time it stops at a finite velocity. The precise scenario involves the parameters
such as the mass difference, difference of the initial velocities and
deviation from the head on collision. An example is shown in Fig.\,6. For more details see
Ref.\ci{PavsicStumbling} where it is concluded:
\begin{quote}
	An example of a higher derivative system is the Pais-Uhlenbeck oscillator. It is a toy model for a higher
	derivative gravity, and for any other system that can be described by a higher derivative theory.
	It is feasible to envisage that there exist or can be fabricated materials that give rise to the
	phenomena, e.g., acoustic waves, whose description can be done in terms of fields satisfying
	a higher derivative action principle. Today, many kinds of so called {\it metamaterials}\index{metamaterials} are known
	with unusual physical properties, including negative index of refraction, negative effective mass
	density\footnote{
		A diametric drive acceleration due to the interaction between positive and
		negative (effective) mass for pulses propagating in a nonlinear mesh lattice
		was obsevred in the experiments performed by Wimmer et al,\,\ci{Wimmer}},
	or exhibiting non linear response to an external influence. This is a fast growing field of
	research with numerous possible applications, and my suggestion here is to explore, amongst others,
	also the possibility of constructing a metematerial behaving in accordance with a variant of
	the action principle (\ref{II2.50}) in which the potential term is bounded from below and above.
\end{quote}

\subsection{Quantum theory}

Let us now illustrate how to quantize a system described by the Lagrangian
\be
L = \frac{1}{2}(\dot x^2  - \dot y^2 ) 
- \frac{1}{2}\omega ^2 (x^2  - y^2 ) .
\lbl{2.15}
\ee
The corresponding Hamiltonian is
\be
H = p_x \dot x + p_y \dot y - L = \frac{1}{2}(p_x^2  - p_y^2 ) 
+ \frac{{\omega ^2 }}{2}(x^2  - y^2 ) ,
\lbl{2.16}
\ee
where
\be
\dotx = \lbrace x,H \rbrace =\frac{\p H}{\p p_x}= p_x~,~~~~~~
{\dot y} = \lbrace y,H \rbrace =\frac{\p H}{\p p_y}= -p_y
\lbl{2.17}
\ee

Upon quantization, $x$, $y$, $p_x$, $p_y$ become operators satisfying the
commutation relations
\be
[x,p_x] = i~,~~~~~~~~[y,p_y]=i
\lbl{2.18}
\ee
\be
[x,y] = 0~,~~~~~~~~[P_x,p_y]=0.
\lbl{2.19}
\ee

Introducing
\bear
&&c_x  = \frac{1}{{\sqrt {2} }}(\sqrt \omega  \,x + 
\frac{i}{{\sqrt \omega  }}\,p_x )\:,\quad c_x^\dg   
= \frac{1}{{\sqrt 2 }}(\sqrt {\omega}  \,x 
- \frac{i}{{\sqrt \omega  }}\,p_x ) \lbl{2.20} \\ 
&&c_y  = \frac{1}{{\sqrt {2} }}(\sqrt {\omega}  \,y 
+ \frac{i}{{\sqrt {\omega}  }}\,p_y )\:,\quad c_y^\dg   
= \frac{1}{{\sqrt {2} }}(\sqrt {\omega}  \,y 
- \frac{i}{{\sqrt {\omega}  }}\,p_y ) \lbl{2.21}
\ear
Eqs.\,(\ref{2.18}),(\ref{2.19}) can be rewritten as
\be
[c_x ,c_x^\dg  ] = 1\:,\qquad [c_y ,c_y^\dg  ] = 1,
\lbl{2.22}
\ee 
\be
[c_x ,c_y ] = 0~,\qquad  [c_x^\dg  ,c_y^\dg  ] = 0 ,
\lbl{2.23} 
\ee
and the Hamiltonian (\ref{2.16} becomes
\be
H = \frac{\omega}{2} \,(c_x^\dg c_x + c_x c_x^\dg - c_y^\dg  c_y- c_y c_y^\dg )
= \omega \,(c_x^\dg  c_x  - c_y^\dg  c_y ).
\lbl{2.24}
\ee
Defining vacuum as
\be
c_x |0\rangle = 0~,~~~~~~~c_y |0\rangle =0.
\lbl{2.25}
\ee
the states with definite energy are created by means of $c_x^\dg$, $c_y^\dg$
according to
\be
   \vac~,~~c_x^\dg \vac~,~~c_x^\dg c_x^\dg\vac~,...,
   c_y^\dg \vac~,~~c_y^\dg c_y^\dg\vac~,...,c_x^\dg c_y^\dg \vac~,...
\lbl{2.7a}
\ee
A generic state,
\be
|mn\rangle =\frac{1}{\sqrt{m! n!}} {c_x^m}^\dg {c_y^n}^\dg \vac,
\ee
satisfies
\be
  H|m n\rangle = \om (m-n) |m n \rangle .
\ee
As in the classical theory, there exist states with positive and negative values of energy.

The norms of the state are all {\it positive}:
\be
   \langle m n| m' n' \rangle = \delta_{m' m'} \delta_{n n'} .
\lbl{2.30}
\ee

This model can be generalized to higher dimensional spaces with signature
$(r,s)$, the Lagrangian and the Hamiltonian being
\be
L = \frac{1}{2}\dot x^a \dot x_a  - \frac{1}{2}\omega ^2 x^a x_a ,
\lbl{2.31}
\ee
\be
H= \frac{1}{2} p^a p_a  + \frac{1}{2}\omega ^2 x^a x_a ,
\lbl{2.32}
\ee

In analogy with (\ref{2.20}),(\ref{2.21}), the annihilation and creation
operators can be defined as
\bear
&&c^a  = \frac{1}{{\sqrt 2 }}\left( {\sqrt \omega  x^a  + 
	\frac{i}{{\sqrt \omega  \,}}p_a } \right) \lbl{2.33}\\ 
&&{c^a}^\dg   = \frac{1}{{\sqrt 2 }}\left( {\sqrt \omega  x^a  
	- \frac{i}{{\sqrt \omega  \,}}p_a } \right), \lbl{2.34} 
\ear
where $p_a = \eta_{ab} p^b$. With such definitions we have
\be
[c^a,{c^b}^\dg]=\delta^{ab}~,~~~~~[c^a,c^b]=[{c^a}^\dg,{c^b}^\dg]=0.
\lbl{2.35}
\ee
\be
c^a |0 \rangle = 0,
\lbl{2.35a}
\ee
\be
H = \frac{1}{2}\,\omega \,(c_a^\dg  c^a  + c^a c_a^\dg  )
=  \omega \,\left (c_a^\dg  c^a + \frac{r}{2} - \frac{s}{2}\right ).
\lbl{2.36}
\ee
and
\be
\langle 0| H |0 \rangle = \frac{\om}{2}(r-s),
\lbl{2.37}
\ee

Alternatively, we can define
\bear
&&a^a  = \frac{1}{2}\left( {\sqrt \omega  \,x^a  
	+ \frac{i}{{\sqrt \omega  }}\,p^a } \right) \lbl{2.38}\\ 
&&{a^a}^\dg   = \frac{1}{2}\left( {\sqrt \omega  \,x^a  
	- \frac{i}{{\sqrt \omega  }}\,p^a } \right). \lbl{2.39} 
\ear
Then
\be
[a^a,a_b^\dg]={\delta^a}_b~, ~~~~~[a^a,{a^b}^\dg]= \eta^{ab}.
\lbl{2.40}
\ee

Concerning vacuum, there are two possible definitions:

{\it Definition} 1. This is the definition usually adopted, namely, $a^a |0 \rangle = 0$. It gives
\be
H = \frac{1}{2}\omega \,(a^{a\dg } a_a  + a_a a^{a\dg } )
=  \omega \,\left (a^{a\dg } a_a  + \frac{r}{2} + \frac{s}{2}\right ) ,
\lbl{2.41}
\ee
Acting on the states created by ${a^a}^\dg$, such Hamiltonian has
{\it positive eigenvalues}. But the norms of the states corresponding to
negative signature are negative, the theory contains ghosts---the unphysical
states.

{\it Definition} 2. Such definition was considered in Refs.\ci{Cangemi,Jackiw},
discussed in \ci{PavsicPseudoHarm}, and reviewed in \ci{PavsicStumbling}.
The operators are split into the positive and negative signature part,
\bear
a^a = (a^\ba,a^{\ul a})~, ~~~~~~~~&&{\bar a}=1,2,...,r~;\nonumber\\
&&{\ul a} = r+1,r+2,...,r+s .
\lbl{2.41a}
\ear 
and vacuum is defined according to
\be
  a^\ba \vac = 0~,~~~~{a^{\ul a}}^\dg \vac =0.
\lbl{2.42}
\ee
This implies that the creation operators are ${a^\ba}^\dg$ and $a^{\ul a}$.
Then Hamiltonian can be written as
\be
H = \omega \,\,\left (a^{\bar a\dg } a_{\bar a}  
+ a_{\ul a} {a^{\ul a}}^\dg  + \frac{r}{2} - \frac{s}{2}\right ) ,
\lbl{2.42a}
\ee
where
\be
  {a^\ba}^\dg a_\ba = \eta_{\ba \bb} {a^\ba}^\dg a^\bb = \delta_{\ba \bb}~,
  ~~~~~,{a_{\ul a}} {a^\ba}^\dg = \eta_{{\ul a} {\ul b}} a^{\ul a} {a^{\ul b}}^\dg
  = - \delta_{{\ul a} {\ul b}} a^{\ul a} {a^{\ul b}}^\dg .
\lbl{2.43}
\ee
Acting on the states created by ${a^\ba}^\dg$, the Hamiltonian (\ref{2.42a})
has positive eigenvalues, whilst acting on the states created by $a^{\ul a}$ it has
negative eigenvalues. Vacuum enery is
\be
   \langle 0|H \vac = \frac{\om}{2} (r-s).
\lbl{2.44}
\ee

According to Woodard\,\ci{Woodard}, the quantization with the vacuum in Definition 2
is the correct one, because it has the correct classical limit and thus
satisfies the correspondence principle. This is not the case with quantization
based on Definition 1, which is therefore incorrect.

We see from (\ref{2.41}) that the ''correct quantization'' gives vanishing
vacuum energy if $r=s$. In Refs.\,\ci{PavsicStumbling} it is then observed:
\begin{quote}
	Analogous happens if we extend the system (\ref{2.31}) to infinite degrees
	of freedom, for instance, to a system of scalar fields, living in a field space
	with neutral metric\,\ci{PavsicPseudoHarm,PavsicSaasFee}. Then the vacuum
	energy is not infinite, but vanishes. Consequently, if such a system is
	coupled to gravity, the notorious cosmological constant problem does not
	arise; the cosmological constant is zero. The observed accelerated cosmological
	expansion, ascribed to dark energy, and possibly due to a small cosmological
	constant, is thus an effect whose explanation
	still remains to be found.
\end{quote}

In the literature is prevailing the acceptance of Definition 1 and the talk is about
negative norm states and ghosts. Among many authors who prefer Definition 2
there is consensus that the systems with negative energies, either classical
or quantum, are problematic anyway, because in physically realistic situations
there are interactions between positive and negative energy degrees of freedom.
As explained in Sec.\,2, that is not true for classical systems in the presence
of physically realistic interactions given by bounded potential--- bounded not
only from below but also from above.

In Refs.\,\ci{PavsicPseudoHarm,PavsicSaasFee} it was shown that also quantum systems that admit
negative energies exhibit unproblematic behavior. This was investigated first
on a toy model described by the Hamiltonian
\be
H = \frac{1}{2}\left( { - \,\frac{\partial }{{\partial x^2 }}
	+ \,\frac{\partial }{{\partial y^2 }}} \right) + V(x,y) .
\lbl{2.45}
\ee
and satisfying the Schr\"odinger equation $i \p \psi/\p t = H \psi$. Numerical
calculation show that nothing strange happened to the wave function
$\psi(t,x,y)$ which evolved so that the probability density $|\psi|^2$ remained
finite and satisfied the normalization condition $\int \dd x \,\dd y |\psi|^2 =1$.
If the initial state is the vacuum, given by\footnote{This can be calculated
by writing the equations $c_x \vac=0$, $c_y \vac$  (see Eq.\,(\ref{2.25})) in 
the $(x,y)$ representation. Such vacuum is not invariant under
the hyperbolic rotations in the $(x,y)$-space. This is not in conflict with covariance of the
theory, because the Hamiltonian (and the corresponding Lagrangian) is invariant.
Therefore, in the new references frame it admits the vacuum solution of the same
form, but expressed in terms of the new coordinates $x'$, $y'$ according to
$\psi'(0) = (2 \pi/\om) {\rm exp}[-\om(x'^2+y'^2)]$. Analogous holds for the excited states.}
\be
\psi_0 = \frac{2 \pi}{\om} {\rm e}^{-\frac{1}{2}\omega (x^2 + y^2)}  ,
\lbl{2.46}
\ee
then at $t>0$ the probability density $|\psi(t,x,y)|^2$ deviates from the
Gaussian shape: it acquires additional picks, which means that the vacuum state has
``decayed'' (see Fig.\,7).
\begin{figure}[h!]
	
	\centerline{\includegraphics[scale=0.70]{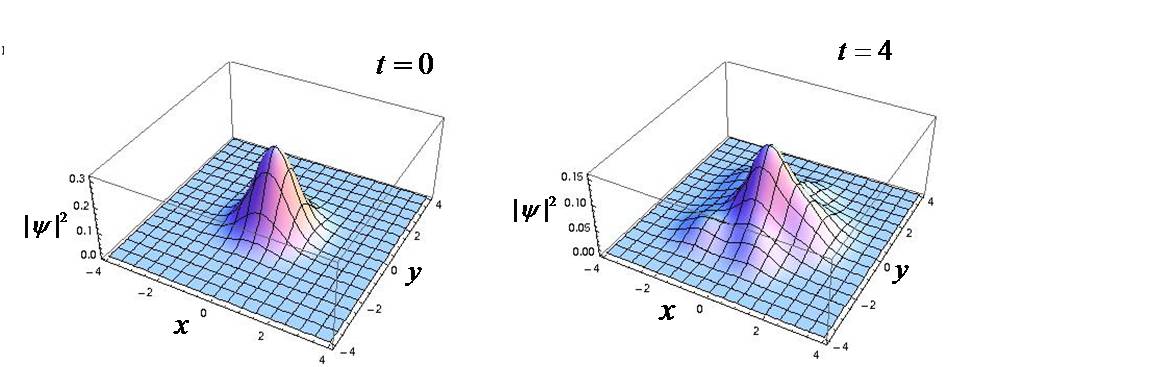}}
	
	\caption{\footnotesize An example of vacuum decay of a 2-dimensional system, given by
	the Hamiltonian (\ref{2.45}). For more details see \ci{PavsicStumbling}.}

\end{figure} 

Extending the consideration to the system with uncountable infinite number of dimensions
we arrive at the action
\be
I = \frac{1}{2}\int {\dd^4 x [g^{\mu \nu } \partial _\mu  \varphi ^a 
	\partial _\nu  \varphi ^b \gam_{ab}  + V(\varphi )]} , 
\lbl{2.47}
\ee
for the system of bosonic fields $\varphi^a$. If the metric $\gam_{ab}$ in the field space
has a signature $(r,s)$, then such system admits negative energies.

Upon quantization the sytem is described by a state vector $|\Psi \rangle$ expanded in terms
of the Fock space basis vectors $ |\bp_1 \bp_2 ... \bp_n \rangle$ that are eigenstates of the
free field Hamiltonian. It satisfies the Schr\"odonger equation
$i \frac{\p|\Psi \rangle}{\p t} = H |\Psi \rangle$.

Because of the presence of negative energy states, the transitions
$\langle \bp_1 \bp_2 ... \bp_n|\Psi (t) \rangle$  can satisfy the energy and momentum
conservation and thus be possible. Therefore, the vacuum decays according to
\be
   \vac \longrightarrow |\Psi (t) \rangle 
   = \sum_{n=0}^\infty \int |\bp_1 \bp_2 ... \bp_n \rangle \dd^3 \bp_1 \dd^3 \bp_2 ...\dd^3 \bp_n
   \langle \bp_1 \bp_2 ... \bp_n|\Psi(t) \rangle,
\lbl{2.49}
\ee
where $\langle \bp_1 \bp_2 ... \bp_n|\Psi(t) \rangle$ is the probability amplitude of
observing at time $t$ the multi particle state $|\bp_1 \bp_2 ... \bp_n \rangle$.
All the probabilities $\langle 0|\Psi\rangle|^2$, $\langle \bp_1|\Psi \rangle|^2$, 
$\langle \bp_1 \bp_2|\Psi \rangle|^2$,...,$\langle \bp_1 \bp_2...\bp_n|\Psi \rangle|^2$
sum to 1:
\be
  \sum_{n=0}^\infty \int \dd^3 \bp_1 \dd^3 \bp_2 ...\dd^3 \bp_n |\langle \bp_1 \bp_2
   ... \bp_n|\Psi(t) \rangle|^2 = 1 .
\lbl{2.50}
\ee

Because the portion of the phase space with a finite number of particles is infinitesimally small in
comparison with the rest of the infinite phase space, at any time $t$ the most probable is
the state with infinite number of particles. In other words\,\ci{PavsicStumbling}
``... the probability that vacuum decays into 2,4,6,8, or any finite number of particles
is infinitely small in comparison with the probability that it decays into infinite numbers
of particles, because such configurations occupy the vast  majority of the phase space.''
From this observation it might be concluded that vacuum instantly decays into infinitely
many particles. This is the majority view in the literature and it is concluded that the
theories allowing for negative energies are not physically viable.

Such conclusion was questioned in Ref.\,\ci{PavsicFirence,PavsicStumbling,Gibbons}.
Namely, when studying the decay rate of a physical system, e.g., an excited nucleus, we
usually do not measure the momenta of the outgoing particles, therefore in the equation
for the transition rate we integrate over them. However, if we {\it do} measure those
momenta, then we do not integrate over them, more precisely, the integration runs
over a narrower portion of the phase space, which reduces the transition rate to a finite value.
Now suppose that the action (\ref{2.47}) describes the universe.\footnote{
In a more realistic model one should include fermions and accompanying gauge fields as well.}
Then $|\Psi(t) \rangle$ of Eq.\,(\ref{2.49}) contains everything in such a universe, including
the observers. For such an observer ${\cal O}$, there is no instantaneous vacuum decay,
because (s)he implicitly or explicitly measures the momenta of the particles in the universe.
Relative to ${\cal O}$, at any time $t$, there is a definite configuration
$|\bp_1 \bp_2...\bp_n \rangle$ (or ``nearly'' definite configuration, given by a narrow wave packet
around $|\bp_1 \bp_2...\bp_n \rangle$). A more detailed description of the arguments along the
lines sketched here can be found in \ci{PavsicFirence,PavsicStumbling} (see also\,\ci{Gibbons}).

\section{Resolving ordering ambiguities}

It is well known
that if gravity action contains terms quadratic in the curvature tensor, then the so obtained
higher derivative gravity upon quantization becomes renormalisable\,\ci{Stelle}.
Having shown that higher derivative
theories are not problematic, a new avenue is opened to quantum gravity. A first step is to
employ the Wheeler-DeWitt equation and then its generalization obtained from quantizing
higher derivative gravity. Unfortunately, we then face with the ordering ambiguity of the
operators in the equation. We will now demonstrate how such problem can be resolved, first for a
scalar, and then for a spinor field. An analogous procedure could be applied to the
Wheeler-DeWitt equation, which is the subject that remains to be investigated in the future.

\subsection{A scalar field}

A typical Hamiltonian in a curved space is given by the expression
\be
H = g^{ab} p_a p_b + V(x) ,
\lbl{3.1}
\ee
where in general the metric $g^{ab}$ depends on position $x\equiv x^a$, $a,b = 1,2,3,...,n$.
Upon quantization the momenta $p_a$ become the operators ${\hat p}_a$ that can be represnted
as $\hp_a = -i \p_a$, where $\p_a \equiv \p/\p x^a$. Because $\hp_a$ acts on the metric
$g^{ab}$, a problem occurs concerning the ordering of operators, namely, what exactly is the
quantum equivalent of the classical expression (\ref{3.1}).

This problem can be resolved\,\ci{PavsicOrder} by means of the geometric calculus based on Clifford
algebras,\ci{Hestenes,Lounesto,Jancewicz,Porteous,Baylis,Lasenby}
in which a vector is given by the expression $A=A^a \gam_a$. Here $A^a$ are vector components and
$\gam_a$ basis vectors, satisfying
\be
  \gam_a \cdot \gam_b \equiv \frac{1}{2} \left ( \gam_a \gam_b + \gam_b \gam_a \right )
  = g_{ab} .
\lbl{3.2}
\ee
In a curved space $\gam_a$ and thus $g_{ab}$ depend on position.

The square of a vector is\footnote{
	The indices are lowered or raised by the metric $g_{ab}$ and its invers $g^{ab}$.}
\be
  A^2 = (A^a \gam_a)(A^b \gam_b) = (\gam^a A_a)(\gam^b A_b) = g_{ab} A^a A^b = g^{ab} A_a A_b.
\lbl{3.3}
\ee
Hence, the classical Hamiltonian (\ref{3.1}) can be written as
\be
  H= p^2 = (\gam^a p_a)(\gam^b p_b) .
\lbl{3.4}
\ee
The corresponding quantum Hamilton operators is then
\be
  {\hat H} = \hp^2 = (\gam^a \hp_a)(\gam^b \hp_b ) = - (\gam^a \p_a)(\gam^b \p_b) ,
\lbl{3.5}
\ee
where $\hp = - i \gam^a \p_a$ is the quantum operators\footnote{
	The expression $- i \gam^a \p_a$ is the {\it definition} of the vector momentum operator.
	The definition with the reversed order, $-i \p_a \gam^a$, makes no sense, because when
	such operators, e.g., acts on a scalar field it gives a non covariant expression
	$-i \p_a (\gam^a \vphi) = -i (\p_a \gam^a) \vphi - i \gam^a \p_a \vphi$.}
corresponding to the classical momentum vectors $p=\gam^a p_a$.

If ${\hat H}$ acts on a scalar field $\vphi (x)$, it gives
\be
  {\hat H} \vphi = -\gam^a \p_a (\gam^b \p_b \vphi)
  = - \gam^a \gam^b \p_a \p_b \vphi - \gam^a \p_b \vphi \p_a \gam^b .
\lbl{3.6}
\ee
In geometric caluclus, $\p_a$ is the derivative whose precise role depends on the object
it acts on. If acting on a scalar field it behaves as the {\it partial derivative}.
If acting on basis vectors it gives the connection:
\be
\p_a \gam^b = - \Gam_{ac}^{\,b} \gam^c .
\lbl{3.7}
\ee
Therefore, if acting on a vector field, $A = A^b \gam_b = A_b \gam^b$, it gives
\be
  \p_a A = 
  \p_a (A^b \gam_b) = \p_a A_b \gam^b + A_b \p_a \gam^b =
  \left (\p_a A_b  - \Gam_{ab}^{\,c} A_c \right ) \gam^b = \DD_a A_b \gam^b ,
  \lbl{3.8}
  \ee
where $\DD_a A_b$ is the {\it covariant derivative} of the vector components.

The action of the Hamilton operator (\ref{3.5}) on a scalar field thus gives
\be
  {\hat H} = \gam^{a} \gam^b \DD_a \p_b \vphi = \gam^a \gam^b \DD_a \DD_b \vphi = g^{ab} \DD_a \DD_b \vphi .
\lbl{3.9}
\ee
There is no ordering ambiguity in this procedure.

\subsection{A spinor field}

Let us also investigate what happens if the derivative acts on a spinor field.
According to Refs.\,\ci{SpinorFock,BudinichP,Giler,Winnberg,PavsicSpinorInverse,BudinichM}, a spinor field is an element
of a minimal left or right ideal of the Clifford algebra $Cl(2n)$. It can be decomposed into basis
spinors $\xi_\alpha$ and the spinor components $\psi^\alpha$:
\be
  \Psi = \psi^\alpha \xi_\alpha~,~~~~~\alpha = 1,2,...,2^{n}.
\lbl{3.10}
\ee

Acting with the derivative on $\xi_\alpha$, we obtain
\be
 \p_a \xi_\al = \Gam_{a \al}^\beta \xi_\beta ,
\lbl{3.11}
\ee
where $\Gam_{a \al}^\beta$ are the coefficients of the spin connection. Therefore,
\be
\p_a \Psi = \left ( \p_a \psi^\al + \Gam_{a \al}^{\,\beta} \psi^\bet \right ) \xi_\al =
\DD_a \psi^\al \xi_\al  ,
\lbl{3.12}
\ee
where $\DD_a \psi^\al$ is the covariant derivative of the spinor components $\psi^\al$.

The Dirac equation can then be written as
\be
   (-\hp + m) \psi = (i \gam^a \p_a + m) \psi = (i \gam^a \DD_a + m) \psi^\al \xi_\al = 0 .
\lbl{3.13}
\ee
Multiplying the latter equation from the left by the reciprocal basis spinors
$\xi^\bet = z^{\bet \gam} \xi_\gam$, where $z^{\bet \gam}$ is the spinors metric\,\ci{PavsicKaluzaLong},
then we obtain the Dirac equation in its usual matrix form
\be
i \left ( {{(\gam^a)}^\bet}_\al + m\,{\delta^\bet}_\al \right ) \psi^\al ,
\lbl{3.14}
\ee
where ${{(\gam^a)}^\bet}_\al = \langle \xi^\bet \gam^a \xi_\al \rangle_S$ are the Dirac
matrices. Here $\langle \rangle_S$ takes the scalar part of the expression\,\ci{Hestenes}.

Multipying the Dirac equation (\ref{3.13}) from the left by $(\hp + m)$, we obtain
\be
  (\hp^2 - m^2) \psi = 0,
\lbl{3.15}
\ee
where $\hp^2 \psi = - \p \p \psi = - \gam^a \p_a (\gam^b \p_b \psi )$. After some
calculation we obtain\,\ci{PavsicStumbling}
\be
  \p \p \Psi = (\gam^a \p_a)(\gam^b \p_b) \Psi = g^{ab} \DD_a \DD_b \psi^\al \xi_\al 
   + \gam^a \wg \gam^b {{R_{ab}}^\al}_\bet \,\psi^\bet \xi_\al ,
\lbl{3.16}
\ee
where $g^{ab} \DD_a \DD_b \psi^\al = g^{ab} \left (\p_a \DD_b \psi^\al - \Gam_{ab}^{\,c} \DD_c \psi^\al
+ \Gam_{a \bet}^{\,\al} \DD_b \psi^\bet \right ) $.

From Eq.\,(\ref{3.16}) we obtain
\be
  \langle \xi^\delta \p \p \Psi \rangle_S =  g^{ab} \DD_a \DD_b \psi^\delta
  + \langle \xi^\delta \gam^a \wg \gam^b \xi_\al \rangle_S {{R_{ab}}^\al}_\bet \,\psi^\bet 
\lbl{3.17}
\ee
where $\langle \xi^\delta \gam^a \wg \gam^b \xi_\al \rangle_S =
\frac{1}{2} {[\gam^a,\gam^b]^\delta}_\al \equiv {{(\sg^{ab})}^\delta}_\al $.

There is no ambiguity about how to take the square of the geometric form of the Dirac
equation (\ref{3.13}): besides the Klein-Gordon part we obtain a coupling term between the
curvature tensor and the spin tensor $\sg^{ab}$.

\section{Strings, branes, braneworlds}

Another approach to quantum gravity and the unification of interactions is based on
strings. A peculiar feature of strings is that though extended they are still singular,
because they are infinitely thin. This means that a string is an idealized description
of an actual physical object. Physical objects are not exactly point-like nor exactly
string-like.
In Refs.\,\ci{CastroChaos,CastroAurilia,Aurilia,CastroPavsicRev,PavsicBook,PavsicKaluzaLong} it was shown how a thick point particle
or a thick string\ci{PavsicSaasFee} can be described by means of the polyvector coordinates,
$X^M \equiv X^{\mu_1 \mu_2 ...\mu_r}$, $r=0,1,2,...,D$ which model an extended object so that
they describe it not only in terms of its center of mass, but also 
in terms of the oriented line, area, volume, 4-volume, ...,$\DD$-volume associated with
the extended object. The coordinates $X^{\mu_1 \mu_2 ...\mu_r}$ are the components
of a polyvector $X = X^M \gam_M$, an element
of the Clifford algebra $Cl(D)$, where  $\gam_M \equiv \gam_{\mu_1 \mu_2 ...\mu_r}$. Such
Clifford algebra is a tangent space to the $D$-dimensional manifold, called Clifford space $C$.

If we start from the 4-dimensional spacetime $M_4$, then there are $2^4=16$ such
coordinates. The configuration space associated with an extended object in $M_4$ has
sixteen dimensions and the signature $(8,8)$\,\ci{PavsicKaluzaLong}.
A realistic string is thus described by sixteen embedding functions $X^M(\tau,\sg)$,
$M=1,2,...,16$, i.e., $X^{\mu_1 \mu_2 ...\mu_r}$, $r=0,1,2,3,4$. It is a string from
the point of view of the 16D Clifford space, but from the point of view of
the 4D spacetime it is a thick string. The quantized string in $C$, whose signature is
neutral, namely $(8,8)$, does not have the problem with the anomalous terms:
the Virasoro algebra closes\\ci{PavsicSaasFee}. A thick string can live in four dimensions. The intricacies
of the usual string theory that require a 26-dimensional (or in the case of the superstring,
10-dimensional) target space do not arise in the theory of thick strings formulated in terms
of polivector coordinates $X^M$, asociated with the 16D Clifford space.

A generalization of strings are branes. The usual Dirac-Nambu-Goto brane, which is described
by the minimal surface action, can be considered as a point particle in an infinite dimensional
space with a particular metric. This can be viewed as a special case of a general
theory\,\ci{PavsicBook,PavsicBrane,PavsicStumbling}
which considers branes as points in the brane space ${\cal M}$, whose metric is dynamical, as
in general relativity. In such theory, besides the Dirac-Nambu-Goto branes, there exist many
other sorts of branes associated with many possible metrics of the brane space ${\cal M}$. 
A partricularly simple is the case of flat brane space, whose metric is the infinite-dimensional
analog of the Minkowski space metric $\eta_{\mu \nu}$.

A brane living in a $D$-dimensonal target space is described by the embedding
functions $X^\mu (\xi^a)$, $\mu=0,1,2,...,D-1$, $a=0,1,2,...,p$. Splitting the parameters
according to $\xi^a = (\tau,\sg^\ba)$, $\ba = 1,2,...,p$, and introducing
the compact notation
\be
  X^{\mu(\sg)} \equiv X^\mu (\tau,\sg)~,~~~~\sg \equiv \sg^\ba ,
\lbl{4.1}
\ee
we can write the brane action in the form of a point particle action\,\ci{PavsicBrane},
\be
I = {\tl \kappa} \int \dd \tau  \left ( \rho_{\mu(\sigma)\nu(\sigma)}  {\dot X}^{\mu (\sigma)}(\tau)
{\dot X}^{\nu (\sigma')}(\tau) \right )^{1/2},
\lbl{4.2}
\ee
where $\rho_{\mu(\sg)\nu(\sg')} \equiv \rho_{\mu \nu}(\sg,\sg')$ is the metric of ${\cal M}$.
If in the latter action we fix a gauge so that ${\dot X}^{\mu(\sg)} {\dot X}_{\mu(\sg)} \equiv
\rho_{\mu(\sigma)\nu(\sigma')} {\dot X}^{\mu (\sigma)} {\dot X}^{\nu (\sigma')}$ is a constant,
say ${\tl k}^2$, then we obtain the Schild action\,\ci{Schild} in infinite dimensions\,\ci{PavsicBrane,PavsicStumbling},
\be
I _{\rm Schild}= \frac{\tl \kappa}{2 {\tl k}} \int \dd \tau \, \rho_{\mu(\sigma)\nu(\sigma')}  {\dot X}^{\mu (\sigma)}(\tau) {\dot X}^{\nu (\sigma')}(\tau) .
\lbl{4.2a}
\ee
The equation of motions derived from the latter action give $(\p/\p \tau) ({\dot X}^{\mu(\sg)} {\dot X}_{\mu(\sg)}) = 0$, which implies that ${\dot X}^{\mu(\sg)} {\dot X}_{\mu(\sg)}$ is an arbitrary constant in $\tau$, determining a gauge.

For a particular choice of metric,
\be
\rho_{\mu(\sigma)\nu(\sigma')} \equiv \rho_{\mu \nu} (\sigma,\sigma') = (-{\bar \gam})\, \eta_{\mu \nu} \, \delta(\sigma-\sigma'),
\lbl{4.3}
\ee
where ${\bar \gam} = {\rm det} \p_\ba X^\mu \p_\bb X_\mu$ is the determinant of the induced metric
on the brane, the action (\ref{4.2}) gives the equations of motion of the Dirac-nambu-Goto brane\,\ci{PavsicBrane,PavsicStumbling}.
This can be directly seen if we plug the metric (\ref{4.3}) into the Schild action (\ref{4.2}). Then we
obtain
\be
 I_{\rm Schild} =\frac{\tl \kappa}{2 \tl k} \int \dd \tau \, \dd^p \sigma \,  (-{\bar \gam}){\dot X}^2,
\lbl{4.4}
\ee
which is the Schild action for the Dirac-Nambu-Goto brane in a gauge in which the determinant of the induced
metric on the brane's worldsheet factorizes as $\gam = {\rm det}\, \p_a X^\mu \p_b X_\mu = {\bar \gam}{\dot X}^2$,
where ${\dot X}^2 \equiv \eta_{\mu \nu} {\dot X}^\mu {\dot X}^\nu$.

Choice of the ${\cal M}$-space metric $\rho_{\mu(\sigma)\nu(\sigma')}$ thus determines a kind of the brane.
The usual brane corresponds to the metric (\ref{4.3}). But in this theory many other metrics are possible\footnote{
In Refs.\,\ci{PavsicBook} a dynamical term for the metric $\rho_{\mu(\sigma)\nu(\sigma')}$ was included into the action, so that like in general relativity, the metric was a solution of the equations of motions, namely, the Einstein equations generalized to the brane space ${\cal M}$.}
and many other kinds of branes.

For the choice
\be
\rho_{\mu(\sigma) \nu(\sigma')} = \eta_{\mu(\sigma) \nu(\sigma')} 
=\eta_ {\mu \nu} \delta (\sigma-\sigma').
\lbl{4.5}
\ee
the action (\ref{4.2}) becomes
\be
I = {\tl \kappa} \int \dd \tau \, \left ( \int \dd^p \, \sigma \, \eta_{\mu \nu} 
{\dot X}^\mu (\tau,\sigma) {\dot X}^\nu (\tau,\sigma) \right )^{1/2} .
\lbl{4.5a}
\ee
which is the infinite dimensional analog of the relativistic point particle action in flat
spacetime, $I= m \int \dd \tau \, (\eta_{\mu\nu} {\dot X}^\mu {\dot X}^\nu )^{1/2}$.

The equations of motion derived from (\ref{4.5a}) are
\be
\frac{\dd}{\dd \tau} \left ( \frac{{\dot X}^\mu (\tau,\sigma)}{{\sqrt{{\dot{\tl X}}^2}}}
\right ) = 0,
\lbl{4.6}
\ee
In a gauge in which ${\dot {\tl X}^2 \equiv {\dot X}^{\mu(\sg)} {\dot X}_{\mu(\sg)} =
\int \dd^p {\dot X}^\mu (\sg} {\dot X}_\mu (\sg) = 1$, the equations of motion
become
\be
{\ddot X}^\mu (\tau,\sigma) = 0,
\lbl{4.7}
\ee
their solution being
\be
X^\mu (\tau,\sigma) = v^\mu (\sigma) \tau + X_0^\mu (\sigma).
\lbl{4.8}
\ee

The latter equation describes a bunch of straight worldlines that form a special kind of
brane's worldsheet: a wordlsheet swept by a ``{\it flat brane}''. The action (\ref{4.5})
with the solution (\ref{4.8}) thus describes a continuum limit of a system of non
interacting point particles that trace straight worldlines. Some examples are illustrated
in Fig.\,8.
\begin{figure}[h!]

	\centerline{\includegraphics[scale=0.50]{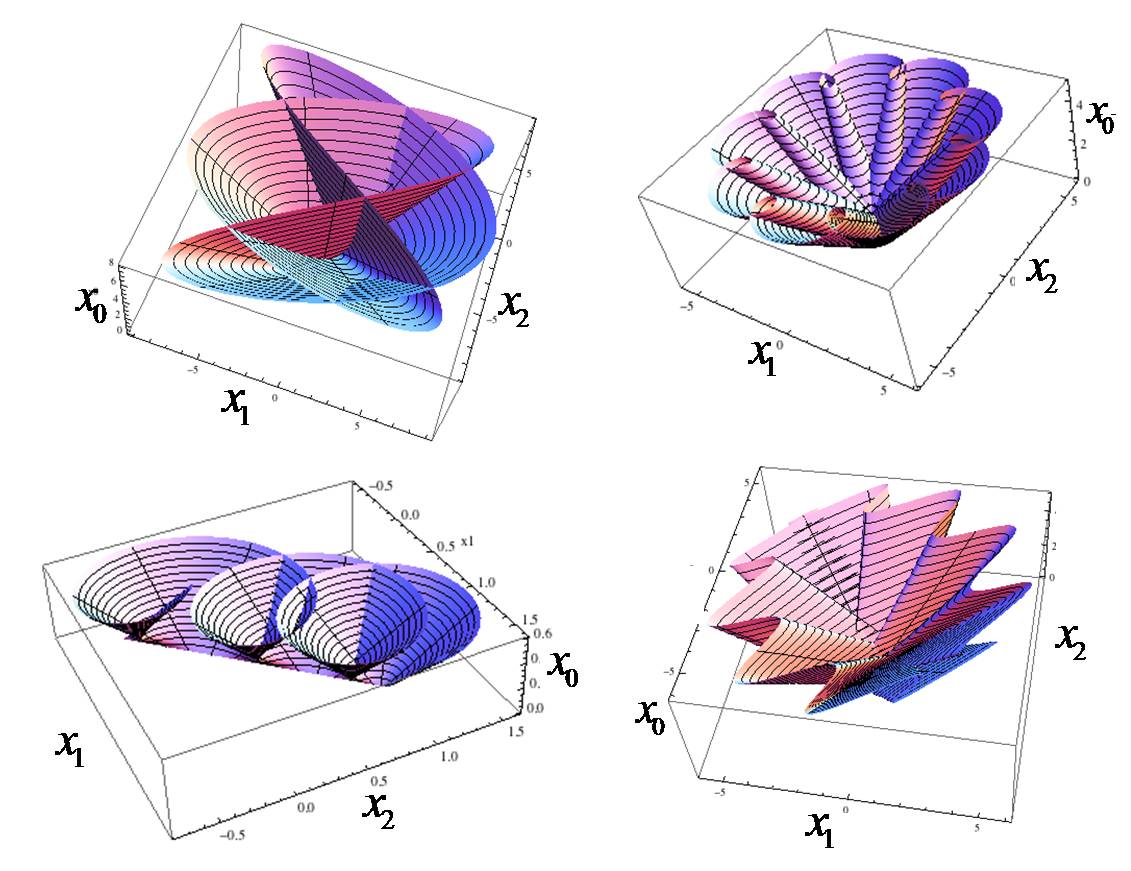}}

	\caption{\footnotesize Examples of flat 1-branes for various choices of initial conditions.}
	\lbl{VIfig2}
\end{figure} 

Upon quantization, each of those free particles within the bunch is described by the
Klein-Gordon equation for a field $\vphi^{(\sg)} \equiv \vphi (\sg,x)$. The action
for whole system is\,\ci{PavsicBrane,PavsicStumbling}
\be
I[\vphi^{(\sigma)}] = \frac{1}{2} \int \dd^D x \, \left ( \p_\mu \vphi^{(\sigma)} \p^\mu \vphi^{(\sigma')}
- m^2 \vphi^{\sigma)} \vphi^{(\sigma')} \right ) s_{(\sigma)(\sigma')},
\lbl{4.9)}
\ee
where $ s_{(\sigma)(\sigma')} \equiv s(\sg,\sg') = \delta_{(\sg)(\sg')} = \delta (\sg-\sg')$
is the metric in the space of the fields $\vphi^{(\sg)}$. In principle, it need not
be such simple metric, it can be a generic metric. A non diagonal metric $s(\sg,\sg')$
determines interactions between the fields. In Ref.,\ci{PavsicBrane,PavsicStumbling} it
was shown that the metric
\be
s(\sg,\sg') = \sqrt{-{\bar \gam} (\sg)}\, \delta^p (\sg-\sg') 
+ \p_\ba \left ( \sqrt{-{\bar \gam} (\sg)} \gam^{\ba \bb} \p_\bb \right )
\delta^{p}(\sg - \sg'),
\lbl{4.10}
\ee
where $\gam \equiv {\rm det} \gam_{\ba \bb}$,
leads to the Dirac-Nambu-Goto brane equation of motion as the expectation value
of the momentum operator in a state with an initially Gaussian wave packet profile.

The limitted space of tis review article does not permit to go into details of the
derivation. Therefore we mention here only the main result, namely that the quantized
brane can be considered as a continuous family of interacting quantum fields.
For a special choice of the interaction, namely, (\ref{4.10}), we obtain the
Dirac-Nambu-Goto brane as the expectation value.

That procedure is thus a novel way to quantization of the brane. One of the much
investigated and promising approaches to quantum gravity are the braneworld scenarios
in which our 4-dimensional spacetime is a brane, more precisely, a 4-dimensional
worldsheet (``worldvolume''), $V_4$, swept by a 3-brane, living in a higher dimensional
embedding space. The induced metric on $V_4$ describes gravity. Quantizing the brane thus
leads to quantum gravity.

\section{Conclusion}

The progress toward quantum gravity is long lasting and difficult. On the way, several
ideas have been rejected and avenues avoided as being physically unviable.
 
First, higher derivative theories and negative energies that they involve have been
put aside, because they seemingly lead to instabilities. But in Ref.\,\ci{Smilga:2004cy,Smilga:2005gb,Smilga:2008pr,Pavsic:2013noa,Pavsic:2016ykq,PavsicStumbling,Gibbons} it was shown that no instabilities occur in physically realistic situations in which the interaction potential
is bounded---not only from below but also from above.

Second, Wheeler DeWitt equation has limited applicability because of the ordering
ambiguities of operators. In Ref.\,\ci{PavsicOrder} a geometric momentum operator
based on Clifford algebra was introduced and shown that in a curved space its square,
acting on a scalar field \,\ci{PavsicOrder} or on a spinor field\,\ci{PavsicStumbling},
is unambiguously defined. A resolution of ordering ambiguities, based on path integrals,
was achieved by Kleinert\,\ci{Kleinert}.

Third, string theory, long considered as very promising, has run into difficulties,
mainly due to the presence of extra dimension and the vast number of possibilities
to compactify them to four dimensions. The problem of dimensional reduction is avoided
by considering a thick string instead of thin string. It was shown that a thick string
formulated in terms of the coordinates of the 16-dimensional Clifford space with
neutral signature can be consistently formulated in four dimensional spacetime\,\ci{PavsicSaasFee}. Clifford space is a truncated configuration space associated with an extended object, described in terms of effective oriented $r$-volumes (lines, areas, 3-volumes, 4-volumes).

Fourth, our universe can be considered as a brane living in a higher dimensional space.
Then the extra dimensions, namely those of the embedding space, need not be compactified.
Unfortunately, quantization of a brane is so difficult that it was not yet successfully
finished. In Ref.\,\ci{PavsicBrane,PavsicStumbling} a wider view was embraced in which
the usual Dirac-Nambu-Goto brane described by the minimal surface action was perceived
as one of many possible different kinds of branes living in an infinite dimensional,
generally curved,
brane space. One such possible brane is the flat brane, whose quantization is
straightforward, because it is just a bunch of non interacting point particles that upon
quantization can be described as a continuous set of free quantum fields. Introducing
interactions between those fields, one obtains as expectation values more complicated
branes. For a particular interfield interaction one obtains the
Dirac-Nambu-Goto brane\ci{PavsicBrane}.

Because of the limited space it was not possible to review here some other
topics relevant for quantum gravity and unification of interactions
presented in Ref.\ci{PavsicStumbling}, e.g, a more detailed exposition of the concept of Clifford
space\ci{CastroChaos,CastroFound,CastroAurilia,Aurilia,PavsicCliff,PavsicBook,PavsicArena,
PavsicMaxwellBrane,CastroPavsicRev},
quantum fields as basis vectors\ci{PavsicSimplectic,PavsicBookLicata,PavsicStumbling}, localization in quantum field theories\,\ci{Fleming1,Fleming2, Karpov,PavsicLocalQFT,PavsicStumbling}, branes as conglomerates of fermionic quantum fields\ci{PavsicBrane,PavsicBookLicata,PavsicStumbling}, and braneworld scenarios\ci{Rubakov,Akama,GibbonsBrane,PavsicEmbed1,PavsicEmbed2,PavsicEmbed3,PavsicEmbed4,Gogberashvili1998,PavsicEmbed5,Gogberashvili2000,PavsicTapia,Brax,PavsicBook},
including braneworlds in Clifford space\,\ci{PavsicSimplectic,PavsicBookLicata}.


\begin{thebibliography}{12}
	\begin{footnotesize}


\bibitem{Smilga:2004cy}
A.~V.~Smilga,
{\it Nucl. Phys.} B \textbf{706}, 598-614 (2005)
doi:10.1016/j.nuclphysb.2004.10.037
[arXiv:hep-th/0407231 [hep-th]].

\bibitem{Smilga:2005gb}
A.~V.~Smilga,
{\it Phys. Lett.} B \textbf{632}, 433-438 (2006)
doi:10.1016/j.physletb.2005.10.014
[arXiv:hep-th/0503213 [hep-th]].

\bibitem{Smilga:2008pr}
A.~V.~Smilga,
SIGMA \textbf{5}, 017 (2009)
doi:10.3842/Sigma.2009.017
[arXiv:0808.0139 [quant-ph]].

\bibitem{Robert:2006nj}
D.~Robert and A.~V.~Smilga,
{\it J. Math. Phys.} \textbf{49}, 042104 (2008)
doi:10.1063/1.2904474
[arXiv:math-ph/0611023 [math-ph]].

\bibitem{Pavsic:2013noa}
M.~Pav\v{s}i\v{c},
{\it Mod. Phys. Lett.} A \textbf{28}, 1350165 (2013)
doi:10.1142/S0217732313501654
[arXiv:1302.5257 [gr-qc]].



\bibitem{Pavsic:2013mja}
M.~Pav\v{s}i\v{c},
{\it Phys. Rev.} D \textbf{87}, no.10, 107502 (2013)
doi:10.1103/PhysRevD.87.107502
[arXiv:1304.1325 [gr-qc]].

\bibitem{Pavsic:2016ykq}
M.~Pav\v{s}i\v{c},
{\it Int. J. Geom. Meth. Mod. Phys.} \textbf{13}, no.09, 1630015 (2016)
doi: 10.1142/ S0219887816300154
[arXiv:1607.06589 [gr-qc]].

\bi{PavsicStumbling}
M.Pav\v si\v c, {\it Stumbling Blocks Against Unification : On Some Persistent Misconceptions in Physics} (World Scientific, 2020).

\bi{PavsicOrder}
M.~Pav\v si\v c,
{\it Class. Quant. Grav.} \textbf{20}, 2697-2714 (2003)
doi:10.1088/0264-9381/20/13/318
[arXiv:gr-qc/0111092 [gr-qc]].

\bi{PavsicSaasFee}
M.~Pav\v si\v c,
{\it Found. Phys.} \textbf{35}, 1617-1642 (2005)
doi:10.1007/s10701-005-6485-x
[arXiv:hep-th/0501222 [hep-th]].


\bi{Rubakov}
 V. A. Rubakov, M. E. Shaposhnikov, 
{\it Physics Letters.} B. 125 (2–3): 136–138 (1983).  
doi:10.1016/0370-2693(83)91253-4 .

\bi{Akama} K. Akama, {\it Lect.\ Notes Phys.} {\bf 176}, 267  (1982) 
[hep-th/0001113].

\bi{GibbonsBrane} G. W. Gibbons and  D. L. Wiltshire,  {\it Nucl. Phys.} {\bf B287}, 717  (1987).

\bi{PavsicEmbed1} M. Pav\v si\v c,
{\it Class. Quant. Grav.} \textbf{2}, 869 (1985)
doi:10.1088/0264-9381/2/6/012
[arXiv:1403.6316 [gr-qc]].

\bi{PavsicEmbed2}
M. Pav\v si\v c,
{\it Phys. Lett.} A \textbf{107}, 66-70 (1985)
doi:10.1016/0375-9601(85)90196-3

\bi{PavsicEmbed3}
M. Pav\v si\v c,
{\it Phys. Lett.} A \textbf{116}, 1-5 (1986)
doi:10.1016/0375-9601(86)90344-0
[arXiv:gr-qc/0101075 [gr-qc]].

\bi{PavsicEmbed4}
M. Pav\v si\v c,
{\it Found. Phys.} \textbf{24}, 1495-1518 (1994)
doi:10.1007/BF02054780

\bi{Gogberashvili1998}
M. Gogberashvili, . "Hierarchy problem in the shell universe 
model", International Journal of Modern Physics D. 11 (10): 1635–
1638(1998) [arXiv:hep-ph/9812296] .

\bi{PavsicEmbed5}
M. Pav\v si\v c,
{\it Phys. Lett.} A \textbf{283}, 8 (2001)
doi:10.1016/S0375-9601(01)00189-X
[arXiv:hep-th/0006184 [hep-th]].

\bi{Gogberashvili2000}
M. Gogberashvili,  "Our world as an expanding shell". 
{Europhysics Letters.} 49 (3): 396–399(2000) [arXiv:hep-ph/9812365]

\bibitem{PavsicTapia}
M.~Pav\v si\v c and V.~Tapia,
``Resource letter on geometrical results for embeddings and branes,''
[arXiv:gr-qc/0010045 [gr-qc]].

\bi{Brax}
Philippe Brax and Carsten van de Bruck,
{\it Class.Quant.Grav.} {\bf 20} R201-R232,2003.

\bi{PavsicBook}
M. Pav\v si\v c,
{\it The Landscape of theoretical physics: A Global view. From point
	particles to the brane world and beyond, in search of a unifying 
	principle} (Kluwer Academic, 2001)
[arXiv:gr-qc/0610061 [gr-qc]].

\bi{PavsicBrane}
M.~Pav\v{s}i\v{c},
{\it Int. J. Mod. Phys.} A \textbf{31}, no.20n21, 1650115 (2016)
doi:10.1142/S0217751X16501153
[arXiv:1603.01405 [hep-th]].

\bibitem{Mostafazadeh:2010yw}
A.~Mostafazadeh,
{\it Phys. Lett.} A \textbf{375}, 93-98 (2010)
doi:10.1016/j.physleta.2010.10.050
[arXiv:1008.4678 [hep-th]].

\bibitem{Banerjee:2013upa}
R.~Banerjee,
``New (Ghost-Free) Formulation of the Pais-Uhlenbeck Oscillator,''
[arXiv:1308.4854 [hep-th]].

\bi{Bolonek}
K. Bolonek, P. Kosínski,
{\it Acta Phys.Polon.} B 36  2115 (2005).

\bi{Sokolov}
E.V. Damaskinsky, M.A. Sokolov,
{\it J.Phys. A} 39  10499 (2006).

\bibitem{Bolonek:2006ir}
K.~Bolonek and P.~Kosinski,
{\it J. Phys. A} \textbf{40}, 11561-11568 (2007)
doi:10.1088/1751-8113/40/38/008
[arXiv:quant-ph/0612091 [quant-ph]].

\bi{Bagarello}
F. Bagarello,
{\it Int.J.Theor.Phys.} {\bf 50}  3241 (2011).

\bibitem{Nucci}
M.~C.~Nucci and P.~G.~L.~Leach,
{\it Phys. Scripta} \textbf{81}, 055003 (2010)
doi:10.1088/0031-8949/81/05/055003
[arXiv:0810.5772 [math-ph]].

\bibitem{Masterov:2015ija}
I.~Masterov,
{\it Nucl.\ Phys.} B {\bf 902}, 95 (2016)
doi:10.1016/j.nuclphysb.2015.11.011
[arXiv:1505.02583 [hep-th]].

\bibitem{Masterov:2016jft}
I.~Masterov,
{\it Nucl.\ Phys.} B {\bf 907}, 495 (2016)
doi:10.1016/j.nuclphysb.2016.04.025
[arXiv:1603.07727 [math-ph]].

\bi{Dector} Aldo D\'ector, Hugo A. Morales-T\'ecotl, Luis F. Urrutia, J. David Vergara,
``Coping with the Pais-Uhlenbeck oscillator's ghosts in a canonical approach",
arXiv:0807.1520 [quant-ph].

\bi{Ghosh} S.~Pramanik and S.~Ghosh,
{\it Mod.\ Phys.\ Lett.} A {\bf 28}, 1350038 (2013)
doi:10.1142/S0217732313500387
[arXiv:1205.3333 [math-ph]].

\bibitem{Kaparulin:2020rqz}
D.~S.~Kaparulin, S.~L.~Lyakhovich and O.~D.~Nosyrev,
{\it Phys. Rev.} D \textbf{101}, no.12, 125004 (2020)
doi:10.1103/PhysRevD.101.125004
[arXiv:2003.10860 [hep-th]].

\bibitem{Abakumova:2019dov}
V.~A.~Abakumova, D.~S.~Kaparulin and S.~L.~Lyakhovich,
{\it AIP Conf. Proc.} \textbf{2163}, no.1, 090001 (2019)
doi:10.1063/1.5130123
[arXiv:1907.08075 [hep-th]].

\bibitem{Abakumova:2019wpn}
V.~A.~Abakumova, D.~S.~Kaparulin and S.~L.~Lyakhovich,
{\it J. Phys. Conf. Ser.} \textbf{1337}, no.1, 012001 (2019)
doi:10.1088/1742-6596/1337/1/012001
[arXiv:1907.02267 [hep-th]].

\bibitem{Kaparulin:2019njc}
D.~S.~Kaparulin,
{\it Symmetry} \textbf{11}, no.5, 642 (2019)
doi:10.3390/sym11050642
[arXiv:1907.03068 [hep-th]].

\bibitem{Abakumova:2019ifi}
V.~A.~Abakumova, D.~S.~Kaparulin and S.~L.~Lyakhovich,
{\it Russ. Phys. J.} \textbf{62}, no.1, 12-22 (2019)
doi:10.1007/s11182-019-01677-0
[arXiv:1905.00263 [hep-th]].

\bibitem{Kaparulin:2018npv}
D.~S.~Kaparulin, I.~Y.~Karataeva and S.~L.~Lyakhovich,
{\it Nucl. Phys. B} \textbf{934}, 634-652 (2018)
doi:10.1016/j.nuclphysb.2018.08.001
[arXiv:1806.00936 [hep-th]].

\bibitem{Abakumova:2017uto}
V.~A.~Abakumova, D.~S.~Kaparulin and S.~L.~Lyakhovich,
{\it Eur. Phys. J.} C \textbf{78}, no.2, 115 (2018)
doi:10.1140/epjc/s10052-018-5601-y
[arXiv:1711.07897 [hep-th]].

\bi{Bondi} H. Bondi,
{\it Rev. Mod. Phys.} {\bf 29}, 423 (1957).

\bi{Shanshan} Shanshan Yao, Xiaoming Zhou and Gengkai Hu, {\it New J. Phys.} {\bf 10}, 043020 (2008).

\bi{Sheng} P. Sheng, X. X. Zhang, Z. Y. Liu  and C. T. Chan, {\it Physica} B {\bf 338}, 201 (2003).

\bi{Mei1} J. Mei, Z. Y. Liu, W. J. Wen and P. Sheng,  {\it Phys. Rev. Lett.} {\bf 96}, 024301 (2006).

\bi{Mei2} J. Mei, Z. Y. Liu, W. J. Wen and P. Sheng, {\it Phys. Rev.} B {\bf 76}, 134205 (2007).

\bi{Choi1} H. Choi  and P. Rudra, {\it Pair Creation Model of the Universe From 
	Positive and Negative Energy}, (2014)
[http://vixra.org/abs/1403.0180].

\bi{Choi2}   H. Choi, {\it Hypothesis of Dark Matter and Dark Energy with Negative Mass},
(2009), 
[http://vixra.org/abs/0907.0015].

\bi{Choi3} H. Choi, {\it On Problems and Solutions of General Relativity}
(Commemoration of the 100th Anniversary of General Relativity)'', (2015), [ http://vixra.org/abs/1511.0240].

\bi{Wimmer}
M. Wimmer, A. Regensburger, C. Bersch, Mohammad-Ali Miri, S. Batz, G. Onishchukov, D. N. Christodoulides and U. Peschel,
{\it Nature Physics} {\bf 9}, 780 (2013).

\bi{Cangemi} D. Cangemi, R. Jackiw and B. Zwiebach, {\it Annals of Physics} {\bf 245}, 408 (1996).

\bi{Jackiw} E. Benedict, R. Jackiw and H. J. Lee, {\it Phys. Rev.} D. {\bf 54}, 6213 (1996).

\bi{PavsicPseudoHarm} M. Pav\v si\v c,
{\it Phys.\ Lett.}\ A {\bf 254}, 119 (1999)
[arXiv:hep-th/9812123].

\bibitem{Woodard} 
R.~P.~Woodard,
{\it Lect.\ Notes Phys.}\  {\bf 720}, 403 (2007)
[arXiv:astro-ph/0601672].

\bi{PavsicFirence}  M.~Pav\v si\v c,
{\it J.\ Phys.\ Conf.\ Ser.}\  {\bf 437}, 012006 (2013)
[arXiv:1210.6820 [hep-th]].

\bi{Gibbons} G.~W.~Gibbons, C.~N.~Pope and S.~Solodukhin,
{\it Phys. Rev.} D \textbf{100}, no.10, 105008 (2019)
doi:10.1103/PhysRevD.100.105008
[arXiv:1907.03791 [hep-th]].

\bi{Stelle} K.~S.~Stelle,
{\it Phys. Rev.} D \textbf{16}, 953-969 (1977)
doi:10.1103/PhysRevD.16.953.

\bi{Hestenes} D. Hestenes,  {\it Space-Time Algebra} (Gordon and Breach 1966);\\
D. Hestenes and  G Sobcyk,
\emph{Clifford Algebra to Geometric Calculus}
(D. Reidel 1984).

\bi{Lounesto} P. Lounesto,  {\it Clifford Algebras and Spinors} (Cambridge
Univ. Press 2001).

\bi{Jancewicz} B. Jancewicz, {\it Multivectors and
	Clifford Algebra in Electrodynamics}, (World Scientific 1988).

\bi{Porteous} R. Porteous,  {\it Clifford Algebras and the Classical Groups}
(Cambridge Univ. Press 1995).

\bi{Baylis}   W. Baylis, {\it Electrodynamics,
	A Modern Geometric Approach} (Birkhauser 1999).

\bi{Lasenby}  A. Lasenby, and C. Doran,  {\it Geometric Algebra for Physicists}
(Cambridge Univ Press 2002). 

\bi{SpinorFock} E. Cartan, {\it Le\c cons sur la th\' eorie des
	spineurs I \& II} (Paris: Hermann, 1938)\\
E. Cartan  {\it The theory of spinors}, English transl. by R.F Streater,
(Paris: Hermann, 1966)\\
C. Chevalley {\it The algebraic theory of spinors}
(New York: Columbia U.P, 1954 )\\
I .M. Benn, R .W. Tucker, {\it An introduction to spinors and geometry with
	appliccations in physics} (Bristol: Hilger, 1987).

\bi{BudinichP} 
P. Budinich  {\it Phys. Rep.} {\bf 137}, 35 (1986);\\
P. Budinich and  A. Trautman  {\it Lett. Math. Phys.} {\bf 11}, 315 (1986).

\bi{Giler} S. Giler,  P. Kosi\' nski, J. Rembieli\' nski and P. Ma\' slanka, 
{\it Acta Phys. Pol.} B {\bf 18}, 713 (1987).

\bi{Winnberg} J. O. Winnberg,  {\it J. Math. Phys.} {\bf 18},  625 (1977).

\bi{PavsicSpinorInverse} M. Pav\v si\v c,
{\it Phys.\ Lett.} B {\bf 692}, 212 (2010)
[arXiv:1005.1500 [hep-th]].

\bi{BudinichM} M. Budinich, {\it J. Math. Phys.}, {\bf 50}, 053514 (2009).\\
M. Budinich,  {\it J. Phys. A}, {\bf 47}, 115201 (2014).\\
M. Budinich, {\it Adv. Appl. Clifford Algebras}, {\bf 25}, 771 (2015).

\bi{PavsicKaluzaLong} M. Pav\v si\v c,
{\it Int.\ J.\ Mod.\ Phys.}\ A {\bf 21}, 5905 (2006)
[arXiv:gr-qc/0507053].

\bi{CastroChaos} C. Castro, {\it Chaos, Solitons and Fractals} 
{\bf 10}, 295 (1999);
{\bf 11}, 1663 (2000);{\bf 12}, 1585 (2001);\\
C. Castro,  {\it Found. Phys.} {\bf 30}, 1301 (2000).

\bi{CastroAurilia} S. Ansoldi, A. Aurilia, C. Castro and E. Spallucci, 
{ Phys. Rev.} D {\bf 64}, 026003 (2001) [hep-th/0105027].

\bi{Aurilia} A. Aurilia, S. Ansoldi and E. Spallucci, {\it 
	Class. Quant. Grav.} {\bf 19},  3207 (2002).

\bi{CastroPavsicRev} C. Castro and M. Pav\v si\v c,
\emph{Prog.\ Phys.} {\bf 1}, 31 (2005).

\bibitem{Schild}  A. Schild, {\it Phys. Rev.} D {\bf 16}, 1722  
(1977).

\bi{Kleinert} H.~Kleinert,
{\it Mod. Phys. Lett.} A \textbf{4}, 2329 (1989)
doi:10.1142/S0217732389002628;
{\it J. Phys. A: Math. Gen.} 29 7619
(1996);\\
H. Kleinert,
{\it Path Integrals in Quantum Mechanics, Statistics and Polymer Physics} (World
Scientific, Singapore, 1995).

\bibitem{PavsicCliff} M. Pav\v si\v c, 
{\it Found.\ Phys.} {\bf 31}, 1185 (2001) [hep-th/0011216].

\bibitem{CastroFound} C. Castro,  {\it Found. Phys.} {\bf 30}, 1301 (2000).

\bi{PavsicArena} M. Pav\v si\v c, {\it Found.\ Phys.} {\bf 33}, 1277  (2003) [gr-qc/0211085]. 

\bi{PavsicMaxwellBrane} M. Pav\v si\v c, \emph{Found.\ Phys.}  {\bf 37}, 1197 (2007)
[hep-th/0605126].

\bi{PavsicSimplectic} M.~Pav\v si\v c,
Adv. Appl. Clifford Algebras \textbf{22}, 449-481 (2012)
doi:10.1007/s00006-011-0314-4
[arXiv:1104.2266 [math-ph]].

\bi{PavsicBookLicata}  M. Pav\v si\v c, ``Quantized Fields \`a 
a Clifford and Unification",
in {\it Beyond Peacefull Coexistence; The Emergence of Space, Time and Quantum"},
Ed. I. Licata (Imperial College Press, 2016).

\bi{Fleming1} G. N. Fleming,
{\it Phys. Rev.} {\bf 137}, B188 (1965).

\bi{Fleming2} G. N. Fleming, ``Lorentz Invariant State Reduction and Localization'',
in {\it Proceedings of the
	Biennial Meeting of the Philosophy of Science Association}, Vol. 1988, Volume Two:
Symposia and Invited Papers (1988), pp. 112--126.

\bi{Karpov} E. Karpov, G. Ordonez, T. Petrosky, I. Prigogine 
and G. Pronko,
{\it Phys. Rev.} A {\bf 62}, 012103 (2000).

\bi{PavsicLocalQFT}
M. Pav\v si\v c,
{\it Adv.\ Appl.\ Clifford Algebras} {\bf 28}, 89 (2018)
doi:10.1007/s00006-018-0904-5
[arXiv:1705.02774 [hep-th]].

\end{footnotesize}
\end{thebibliography}
\end{document}